\documentclass[a4paper,fleqn,usenatbib]{mnras}
\pdfoutput=1

\usepackage[pdftex]{graphicx,xcolor}
\usepackage[T1]{fontenc}
\usepackage[utf8]{inputenc}
\usepackage{lmodern}
\usepackage{float}
\usepackage{caption}

\definecolor{urlblue}{rgb}{0,0,0.9}
\definecolor{linkgreen}{rgb}{0,0.45,0}
\definecolor{linkorange}{rgb}{0.7,0.1,0.0}

\usepackage{amsmath, amssymb}
\usepackage{cuted}
\usepackage[english]{babel}
\usepackage{enumerate}
\usepackage[normalem]{ulem}
\usepackage{enumitem}
\usepackage{multirow}
\usepackage{booktabs}
\usepackage{textcase}
\usepackage{makecell}
\usepackage{xspace}

\usepackage[table]{xcolor}  
\usepackage{colortbl}       
\usepackage{array}          

\definecolor{uBand}{HTML}{56B4E9}   
\definecolor{gBand}{HTML}{009E73}   
\definecolor{rBand}{HTML}{D55E00}   
\definecolor{iBand}{HTML}{E69F00}   
\definecolor{zBand}{HTML}{CC79A7}   
\definecolor{yBand}{HTML}{999999}   

\usepackage{tikz}
\usetikzlibrary{shapes.geometric, arrows, positioning, fit, arrows.meta}
\tikzstyle{startstop} = [rectangle, rounded corners, minimum width=2.cm, minimum height=0.8cm, text centered, draw=black]
\tikzstyle{process} = [rectangle, minimum width=2.cm, minimum height=0.8cm, text centered, draw=black]
\tikzstyle{decision} = [diamond, minimum width=2.cm, minimum height=0.8cm, text centered, draw=black]
\tikzstyle{arrow} = [thick,->,>=stealth]

\usepackage{fourier}
\DeclareMathAlphabet{\mathcal}{OMS}{cmsy}{m}{n}
\SetMathAlphabet{\mathcal}{bold}{OMS}{cmsy}{b}{n}

\definecolor{valecol}{rgb}{0,0.5, 1.}

\definecolor{rmcol}{rgb}{1.,0.5, 0.}

\definecolor{rikcol}{rgb}{1,0,0}

\def\d{{\rm d}}

\newcommand{\Mhalo}{{\ifmmode{M_{\rm halo}}\else{$M_{\rm halo}$}\fi}}

\newcommand{\sky}{439.8\xspace}

\graphicspath{{./figs/}}

\AtBeginDocument{\hypersetup{
linkcolor=linkgreen,
citecolor=linkorange,
urlcolor=urlblue}}

\title[GW and EM detection in the context of the \texttt{CosmoDC2} LSST synthetic catalog]{Gravitational-wave and electromagnetic detections in the context of the \texttt{CosmoDC2} LSST synthetic catalog}

\author[Menote et al.]{
Ranier Menote,$^{1,2}$\thanks{E-mail: \href{mailto:ranier_m@hotmail.com}{ranier\_m@hotmail.com}}
Valerio Marra,$^{2,3,4,5}$
Riccardo Sturani,$^{6,7}$
Felipe Andrade-Oliveira,$^{8}$
Clécio R.\ Bom,$^{9}$\newauthor
and The LSST Dark Energy Science Collaboration
\\
$^{1}$PPGCosmo, Universidade Federal do Espírito Santo, 29075-910 Vitória, ES, Brazil\\
$^{2}$Laboratório Interinstitucional de e-Astronomia – LIneA, 20921-400 Rio de Janeiro, RJ, Brazil\\
$^{3}$Departamento de Física, Universidade Federal do Espírito Santo, 29075-910 Vitória, ES, Brazil\\
$^{4}$INAF – Osservatorio Astronomico di Trieste, via Tiepolo 11, 34131 Trieste, Italy\\
$^{5}$IFPU – Institute for Fundamental Physics of the Universe, via Beirut 2, 34151 Trieste, Italy\\
$^{6}$Instituto de Física Teórica, Universidade Estadual Paulista, 01140-070 São Paulo, SP, Brazil\\
$^{7}$ICTP South American Institute for Fundamental Research, 01140-070 São Paulo, SP, Brazil\\
$^{8}$Physik Institut, University of Zurich, Winterthurerstrasse 190, 8057 Zürich, Switzerland\\
$^{9}$Centro Brasileiro de Pesquisas Físicas, 22290-180 Rio de Janeiro, RJ, Brazil
}

\date{Accepted XXX. Received YYY; in original form ZZZ}

\pubyear{202X}

\begin{document}

\label{firstpage}

\pagerange{\pageref{firstpage}--\pageref{lastpage}}

\maketitle

\begin{abstract} 
Third-generation (3G) gravitational-wave (GW) detectors will enable precision cosmology with dark and bright sirens, motivating realistic simulations that jointly model GW sources, electromagnetic (EM) counterparts, and host galaxies. We present a self-consistent framework that embeds binary compact object mergers within galaxies from the \texttt{CosmoDC2} LSST synthetic catalog. We model ab initio the mass, spin, and redshift distributions of binary black holes (BBH), binary neutron stars (BNS), and black hole-neutron star (BHNS) systems using astrophysically motivated prescriptions, and sample host galaxies using redshift- and mass-dependent merger-rate templates calibrated on population-synthesis results.
We generate GW signals for multiple waveform approximants and detector networks spanning second-generation (LIGO, Virgo, KAGRA) and third-generation observatories (Einstein Telescope, Cosmic Explorer), including realistic duty cycles. Parameter covariances are estimated with Fisher-matrix methods, while EM counterparts are modeled via kilonova prescriptions to produce synthetic LSST-band photometry.
We publicly release \href{https://github.com/LSSTDESC/CosmoDC2_BCO}{\protect\texttt{CosmoDC2\_BCO}}, providing intrinsic and extrinsic source parameters, signal-to-noise ratios, parameter uncertainties, sky areas, and kilonova magnitudes. We find that 3G networks dramatically increase detection rates and improve parameter estimation, and that retaining 2G detectors alongside 3G facilities can significantly enhance sky localization and distance precision, particularly for BNS. Under a simplified Target-of-Opportunity strategy, an LSST-like survey paired with a CE+ET+LVK network at 70\% duty cycle could detect about 5\,000 GW-associated kilonovae over 10 years on a 16\,000~deg$^{2}$ footprint. These forecasts depend on merger-rate and kilonova-luminosity assumptions and should be interpreted accordingly.
\end{abstract}


\section{Introduction}

Gravitational-wave (GW) astronomy has progressed rapidly in the decade following the first direct detection of a binary black hole (BBH) merger in 2015 \citep{Abbott:2016blz}, offering transformative insights into binary compact object populations and enabling novel probes of the universe. The upcoming third-generation (3G) detectors, such as the Einstein Telescope \citep[ET,][]{Maggiore:2019uih} and Cosmic Explorer \citep[CE,][]{Reitze:2019iox}, are expected to significantly increase the number of detected events and extend the high end of the signal-to-noise ratio distribution. This will enable precise measurements of cosmological parameters, including the Hubble constant and the time evolution of dark energy, through independent dark \citep{DES:2019ccw, Bom:2024afj,LIGOScientific:2025jau} and bright \citep{Palmese:2023beh, Abbott:2017xzu} siren observations, with the potential to shed light on the unresolved Hubble tension \citep{Abdalla:2022yfr,CosmoVerseNetwork:2025alb} and the nature of dynamical dark energy \citep{DESI:2025zgx}.

In this work, we investigate the scientific potential of future GW observations in combination with the Legacy Survey of Space and Time (LSST), which will be conducted at the Vera C.\ Rubin Observatory\footnote{\url{https://rubinobservatory.org}} starting in late 2025. We simulate binary compact object (BCO) mergers—including BBH, binary neutron star (BNS), and black hole–neutron star (BHNS) systems—using realistic astrophysical and cosmological models.
Host galaxies are consistently drawn from the \texttt{CosmoDC2} synthetic galaxy catalog \citep{LSSTDarkEnergyScience:2019hkz}, developed for precision dark energy science with LSST \citep{LSSTDarkEnergyScience:2018jkl,LSSTDarkEnergyScience:2020oya}, ensuring a physically motivated association between GW sources and their environments.
This enables robust forecasts for both dark and bright siren cosmology, capturing correlations induced by the large-scale structure, including galaxy clustering and peculiar velocities. By incorporating LSST-like observational properties, such as photometric uncertainties and inferred stellar masses, we establish a coherent and self-consistent framework for multi-messenger cosmology.

The full set of input, output, and value-added catalogs produced by the pipeline illustrated in Figure~\ref{fig:flowchart} is publicly available at \href{https://github.com/LSSTDESC/CosmoDC2_BCO}{\texttt{github.com/LSSTDESC/CosmoDC2\_BCO}}. The busy reader may refer to Table~\ref{tab:data-prod}, which summarizes the structure of the data products. With this paper, we release Version~1 of \texttt{CosmoDC2\_BCO}~\citep{menote2025-zenodo}, to be continuously updated as improved astrophysical models, cosmological assumptions, and inference tools become available.

Here, in addition to a detailed description of the methodology, we focus on the presentation of the gravitational-wave catalog and its electromagnetic follow-up. The full exploitation of the synergies among galaxy, gravitational-wave, and electromagnetic data for cosmological inference is deferred to future work. We simulate a range of detector network configurations, waveform models, and observational scenarios, including different duty cycles, to quantify their impact on detection rates and parameter estimation. We present comprehensive forecasts for the precision of astrophysical and cosmological parameters attainable with various combinations of second-generation (2G) and 3G detector networks, explicitly analyzing the added value of retaining 2G observatories in conjunction with 3G infrastructure. We also evaluate the detectability of kilonova counterparts to neutron star mergers with an LSST-like survey, highlighting the prospects for multimessenger observations. Our results provide strategic insights for optimizing observational design and guiding future developments in gravitational-wave and time-domain astronomy.

This paper is organized as follows. In Section~\ref{section_pipeline}, we detail the methodology used to generate simulated gravitational-wave catalogs, including the modeling of host galaxies, binary properties, waveform approximants, detector networks, inference procedures, and EM counterparts. The pipeline is summarized in Figure~\ref{fig:flowchart}.
Section~\ref{results} presents the results, including network sensitivity, detection counts, parameter uncertainties, duty cycle effects, and electromagnetic counterparts. The Appendices present complementary analyses. We conclude in Section~\ref{conclusions}.

\section{Methodology}\label{section_pipeline}

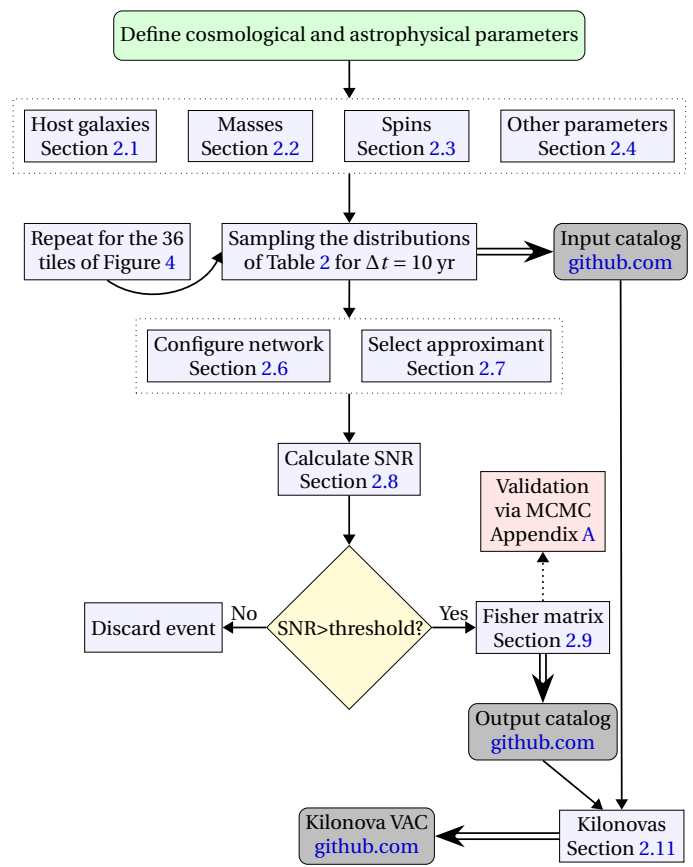
\begin{figure}
\centering
\resizebox{\columnwidth}{!}{  
\begin{tikzpicture}[node distance=.5cm]

\node (start) [startstop, fill=green!15] {Define cosmological and astrophysical parameters};

\node (masses) [process, below=0.8cm of start, xshift=-1.37cm, align=center, fill=blue!5] {Masses \\ Section~\ref{sec:mass-dist}};
\node (sky) [process, left=of masses, align=center, fill=blue!5] {Host galaxies \\ Section~\ref{HG_models_sec}};
\node (spins) [process, right=of masses, align=center, fill=blue!5] {Spins \\ Section~\ref{spins_sec}};
\node (angles) [process, right=of spins, align=center, fill=blue!5] {Other parameters \\ Section~\ref{Further_para_sec}};

\node (dottedbox) [draw, dotted, fit={(sky) (masses) (spins) (angles)}, inner sep=5pt] {};

\node (event) [process, below=0.8cm of dottedbox, align=center, fill=blue!5] {Sampling the distributions \\ of Table~\ref{tab:parameters_priors} for  $\Delta t=10$ yr};

\node (tiles) [process, left=of event, align=center, fill=blue!5] {Repeat for the 36  \\ tiles of Figure~\ref{fig:tiling}};

\node (input) [startstop, right=1.23cm of event, align=center, fill=lightgray] {Input catalog\\ \href{https://github.com/LSSTDESC/CosmoDC2_BCO}{github.com}};

\node (detector) [process, below=0.8cm of event, align=center, xshift=-1.78cm, fill=blue!5] {Configure  network  \\ Section~\ref{networks}};
\node (waveform) [process, right=of detector, align=center, fill=blue!5] {Select approximant \\ Section~\ref{wave-models}};

\node (dottedbox2) [draw, dotted, fit={(waveform) (detector)}, inner sep=5pt] {};

\node (snr) [process, below=0.8cm of dottedbox2, align=center, fill=blue!5] {Calculate SNR \\ Section~\ref{sec:snr}};

\node (decision) [decision, below=0.8cm of snr, text width=2.cm, align=center, fill=yellow!20] {\!\!\!SNR>threshold?};

\node (computeFisher) [process, right=.7cm of decision, align=center, fill=blue!5] {Fisher matrix \\ Section~\ref{inference}};
\node (mcmc) [process, above=0.8cm of computeFisher, align=center, fill=red!10] {Validation \\ via MCMC \\ Appendix~\ref{ap:validation}};
\node (output) [startstop, below=0.8cm of computeFisher, align=center, fill=lightgray] {Output catalog\\ \href{https://github.com/LSSTDESC/CosmoDC2_BCO}{github.com}};

\node (kn) [process, below right=.8cm and -.8cm of output, align=center, fill=blue!5] {Kilonovas \\ Section~\ref{sec:kn}};
\node (vac) [startstop, left=2cm of kn, align=center, fill=lightgray] {Kilonova VAC \\ \href{https://github.com/LSSTDESC/CosmoDC2_BCO}{github.com}};

\node (discard) [process, left=.7cm of decision, fill=blue!5] {Discard event};

\draw [thick, -{Triangle[scale=1.]}] (start.south) -- (dottedbox.north);
\draw [thick, -{Triangle[scale=1.]}] (dottedbox.south) -- (event.north);
\draw [thick,  double distance=2pt, -{Stealth[scale=1.5]}]  (event.east) -- (input.west);
\draw [thick, -{Triangle[scale=1.]}] (event.south) -- (dottedbox2.north);
\draw [thick, -{Triangle[scale=1.]}, bend right=50] (tiles.south) to (event.west);
\draw [thick, -{Triangle[scale=1.]}] (dottedbox2.south) -- (snr.north);
\draw [thick, -{Triangle[scale=1.]}] (snr.south) -- (decision.north);

\draw [thick, -{Triangle[scale=1.]}] (decision.east) -- (computeFisher.west) node[midway,above] {Yes};
\draw [thick, -{Triangle[scale=1.]}] (decision.west) -- (discard.east) node[midway,above] {No};
\draw [thick,  double distance=2pt, -{Stealth[scale=1.5]}]  (computeFisher.south) -- (output.north);
\draw [thick, -{Triangle[scale=1.]}, dotted] (computeFisher.north) -- (mcmc.south);

\draw [thick, -{Triangle[scale=1.]}] (input.south) -- (kn.north);
\draw [thick, -{Triangle[scale=1.]}] (output.south) -- ([xshift=-3mm]kn.north);
\draw [thick,  double distance=2pt, -{Stealth[scale=1.5]}]  (kn.west) -- (vac.east);

\end{tikzpicture}
}
\caption{Workflow for generating  gravitational-wave catalogs.}
\label{fig:flowchart}
\end{figure}

Accurate and unbiased cosmological and astrophysical inferences from simulated BCO mergers require careful selection of input parameters, as they critically affect both astrophysical predictions—such as binary properties—and cosmological measurements, particularly the inference of the Hubble constant from bright sirens, where precise redshift estimates are essential.  
To ensure realism and robustness, we adopt physically motivated parameter distributions in constructing our simulated event catalogs.  
Figure~\ref{fig:flowchart} summarizes the ab initio pipeline implemented in this work, which is described in detail throughout this section.

\subsection{Host Galaxies Population}
\label{HG_models_sec}

The first step in simulating our catalogs is to sample the redshifts (luminosity distances) and sky positions (right ascension and declination) of the host galaxies. To this end, we adopted the novel template model described in this section to fit the merger rate density derived from the simulations of \citet{Santoliquido:2022kyu}. That study characterized the host galaxies of BCO mergers using observational quantities such as stellar mass, star formation rate (SFR), and average metallicity.
Given the properties of the \texttt{CosmoDC2} catalog described in Section~\ref{Cosmo_DC2_sec}, we chose the stellar mass $M$ as the primary variable to model the merger rate, as it was identified by \citet{Artale:2020swx} as the most reliable predictor of merger occurrence among galaxy properties.

The merger rate template model that we adopt incorporates am overall redshift dependency inspired by the functional form given by \citet{Madau:2014bja}:
\begin{align}
\psi(z) = N_{\mathcal{R}} \frac{(1+z)^A}{1 + \left(\frac{1+z}{C}\right)^B} \,,
\end{align}
which is widely adopted for modeling star formation rates across different epochs of the universe.
This choice is motivated by the anticipated correlation between star formation activity and compact binary mergers.%
\footnote{We do not model the merger rate as a stochastically delayed star formation rate, as in \citet{Leandro:2021qlc}. Instead, we adopt the star formation rate template with free parameters, multiplied by a phenomenological function of mass that more directly captures the typical delay between star formation and binary compact object mergers.}
Here, as explained below, the parameters $N_{\mathcal{R}}, A, B, C$ will be fitted to the simulations.

As shown by \citet{Santoliquido:2022kyu}, the merger rate of BNS exhibits a different dependence on stellar mass compared to that of BBH and BHNS. To accurately account for these differences, we adopt separate parameterizations of the merger rate model for BNS and for BBH/BHNS systems, as detailed in the following two sections.

\begin{figure*}
\centering 
\includegraphics[trim={0 0 0 0}, clip, width= 1 \textwidth]{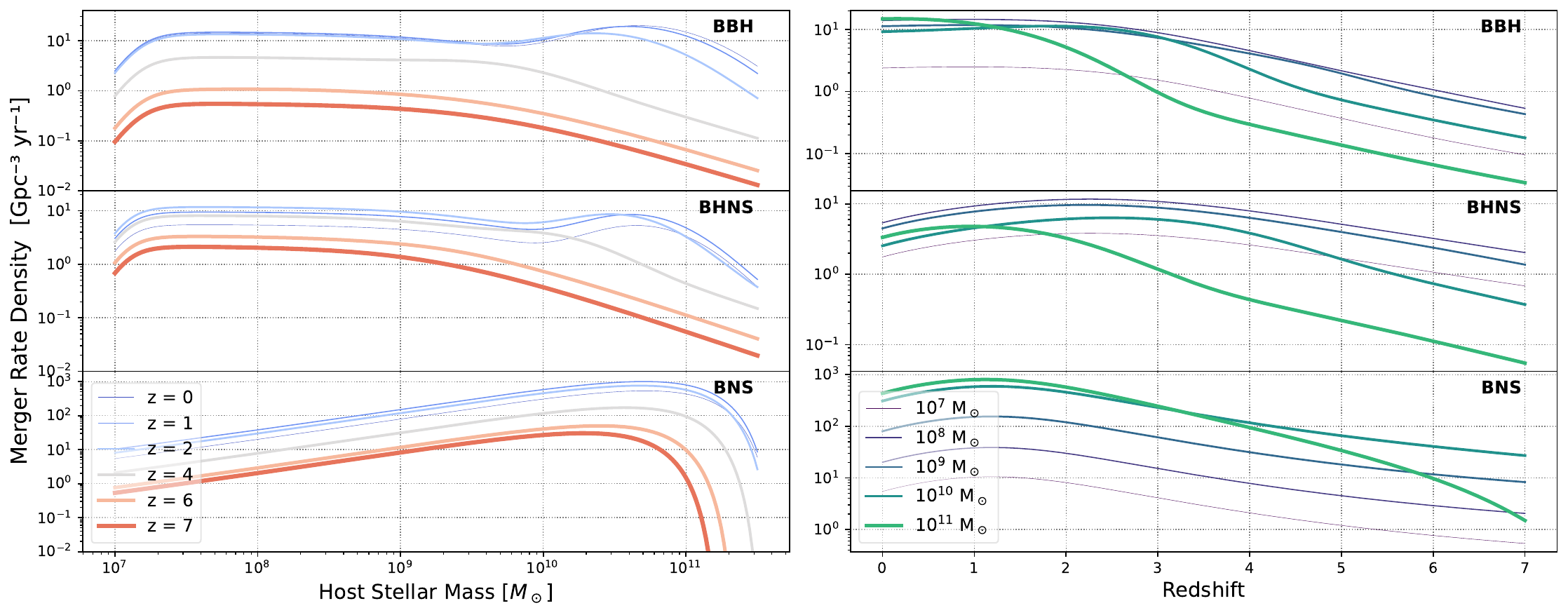}
\caption{Merger rate density as a function of mass at fixed redshifts (left) and as a function of redshift at fixed masses (right) for BBH (top), BHNS (middle), and BNS (bottom). See Section~\ref{HG_models_sec} for details.}
\label{Partial_MRD}
\end{figure*}

\subsubsection{BBH and BHNS merger rates}\label{sec:bbhmerger}

\begin{table}
\centering
\caption{Best-fit values of the parameters of the templates of Section~\ref{HG_models_sec} when fitting to the simulations by \citet{Santoliquido:2022kyu}.\label{tab:parameters}}
\setlength{\tabcolsep}{6pt}
\renewcommand{\arraystretch}{1.2}
\begin{tabular}{|c|c||c|c|}
\hline
\textbf{BBH/BHNS} & \textbf{Value} & \textbf{BNS} & \textbf{Value}\\ \hline
$N_{\mathcal{R}}$ [Gpc$^{-3}$ yr$^{-1}$] & 7.27 / 2.78  & $N_{\mathcal{R}}$ & 541.7 \\ \hline
$A $ & 0.09 / 0.82 & $A $ & 1.17 \\ \hline
$B $ & 5.28 / 4.55 & $B $ & 4.95 \\ \hline
$C $ & 4.19 / 4.55 & $C $ & 2.73 \\ \hline
$K_1 $ & 4.73 / 5.38 & $\nu_m$ & 11.68 \\ \hline
$K_2 $ & 0.99 / 1.03 & $K$ & $7.8 \!\times\! 10^{-4}$ \\ \hline
$M_1 $ & $1.46 / 1.17 \!\times\! 10^7$ & $L_c$ & -0.029  \\ \hline
$M_2 $ & $4.30 / 4.95 \!\times\! 10^{9}$ & $\sigma_{\mathcal{_R}}$ & 0.22 \\ \hline
$C_1 $ & -$3.4 \!\times\! 10^{-4}  /  1.75 \!\times\! 10^{-2}$ & - & - \\ \hline
$C_2 $ & -$9.53 \!\times\! 10^{-5} / \text{-}1.19 \!\times\! 10^{-2}$ & - & - \\ \hline
$G $ & 2.52 / 1.67 & - & - \\ \hline
$G_c $ & 10.69 / 10.68 & - & - \\ \hline
$G_{\rm sig}  $ & 0.41 / 0.34 & - & - \\ \hline
$G_d $ & 0.10 / 0.10& - & - \\ \hline
$ \nu $ & 0.07 / 0.04& - & - \\ \hline
\end{tabular}
\end{table}

Considering now the case of BBH and BHNS, to account for the stellar mass dependence of the host galaxies—where more massive galaxies may indicate increased star formation and, consequently, more mergers—, we introduce two functions: a Heaviside step function $\mathcal{H}$ and a Gaussian distribution $\mathcal{G}$, centered on more massive galaxies.
The merger rate density per unit of \(\log_{10}(M)\) in the source frame per comoving volume is then described by the following expression:\footnote{This  is denoted as \(\d \mathcal{R}/ \d \log_{10}M\) in \citet{Santoliquido:2022kyu}.}
\begin{align}
\label{SourceMRD}
&\mathcal{R}(z, M, \theta_{\rm BBH/BHNS}) = \psi \times (\mathcal{H} + \mathcal{G}) \,.
\end{align}
Here, $M$ stands for stellar mass and the parameters governing the redshift evolution and the stellar mass distribution of the host galaxy are collectively denoted by the vector
\begin{align} \label{parBH}
\theta \!=\! \{ N_{\mathcal R}, A, B, C, K_1, K_2, M_1, M_2, C_1, C_2, G, G_c, G_{\text{sig}}, G_d, \nu \} . 
\end{align}
The Heaviside step and Gaussian components are defined as:
\begin{align}
\label{HS_fun}
\mathcal{H} &= \tanh \big (K_1[\log_{10}{M}  - \log_{10}{M_{\rm 1}}] \big) \nonumber \\ 
&\quad - \tanh \big (K_2[\log_{10}{M} - \log_{10}{M_{\rm 2}} - C_1 \cdot z - C_2 \cdot z^2] \big ) \,, \\
 \mathcal{G} &= G \cdot \exp\left(-0.5 \left(\frac{\log_{10}(M) - (G_c - \nu z^2)}{G_{\text{sig}}}\right)^2 - G_d \cdot z^2\right) \,.
\end{align}
Note that the limits in the Heaviside function for massive host galaxies depend on redshift, such that mergers at higher redshifts are concentrated in galaxies with lower stellar mass. Similarly, the center and amplitude of the Gaussian peak for host galaxy stellar mass depend on redshift, damping its contribution at higher redshifts.

\citet{Santoliquido:2022kyu} generates synthetic galaxy populations from redshift-dependent observational scaling relations linking stellar mass, star formation rate, and metallicity, thereby encoding galaxy evolution. These galaxies are then populated with binary compact objects whose merger delay times are obtained from binary population-synthesis calculations that include the common-envelope phase, parameterized by \(\alpha\). This approach couples the build-up of galaxy populations to the formation and merger of compact binaries.

During the common-envelope phase, the orbital energy released as the components spiral in is compared to the binding energy of the stellar envelope. The parameter \(\alpha\) quantifies the efficiency with which this orbital energy unbinds the envelope \citep{Hurley:2002rf}. Smaller \(\alpha\) implies inefficient ejection, forcing a deeper inspiral to supply sufficient energy; if the system survives, it exits the phase in a much tighter orbit and with a shorter delay time. Larger \(\alpha\) corresponds to more efficient ejection, requiring less orbital shrinkage and yielding wider post-common-envelope binaries with longer delay times. Hence \(\alpha\) shapes the delay-time distribution and, consequently, the merger-rate density by regulating both the survival probability through the common-envelope phase and the timescale to coalescence. In addition, the overall normalization depends on the cosmic star-formation history encoded in the galaxy model. \citet{Santoliquido:2022kyu} show that alternative prescriptions for the stellar mass--star formation rate relation can shift the predicted merger-rate normalization by factors of \(\sim 3\)--\(4\).
In this work, we adopt the merger-rate density from the simulation assuming the Fundamental Metallicity Relation (FMR) for metallicity and the \(\alpha_5\) binary model, as labeled in \citet{Santoliquido:2022kyu}, since it best matches the observational constraint discussed in Eq.~\eqref{z02bbh}.

Table~\ref{tab:parameters} summarizes the best-fit parameters obtained by fitting the BBH and BHNS merger rate templates to these simulations across the redshift interval \(0<z<7\).%
\footnote{\href{https://gitlab.com/Filippo.santoliquido/galaxy_rate_open}{\tt gitlab.com/Filippo.santoliquido/galaxy\_rate\_open}}
Using this fiducial parameterization, we obtain a local merger rate density at \( z=0.2 \) given by:
\begin{align} \label{z02bbh}
\int \mathcal{R}(z=0.2,M, \theta_{\rm BBH})\,\mathrm{d}\log_{10}M = 55.3~\text{Gpc}^{-3}\,\text{yr}^{-1}\,.
\end{align}
This value is somewhat higher than the empirical \(90\%\) credible interval \([17.9,\,44]~\text{Gpc}^{-3}\,\text{yr}^{-1}\) reported by \citet{KAGRA:2021duu}, but remains plausible given the substantial statistical and systematic uncertainties affecting both observational inference and astrophysical modeling.

In particular, the empirically inferred rate depends on population and selection assumptions, while forward models are sensitive to the cosmic star-formation history, binary-evolution parameters (notably the common-envelope efficiency \(\alpha\)), and metallicity evolution; hence our fiducial choice is well within the expected modeling uncertainty. It is also important to note that the credible interval \([17.9,\,44]~\mathrm{Gpc}^{-3}\,\mathrm{yr}^{-1}\) reported by \citet{KAGRA:2021duu} is derived from a restricted set of high-significance BBH events. When lower-significance candidates (e.g.\ with \(p_{\rm astro}>0.5\)) up to the end of O3 are included, \citet{KAGRA:2021duu} find a broader \(90\%\) credible interval, \([24.8,\,63.6]~\mathrm{Gpc}^{-3}\,\mathrm{yr}^{-1}\), which encompasses our fiducial normalization. Moreover, the predicted local rate is highly sensitive to assumptions about the cosmic star-formation and metallicity histories, as well as the modeling of binary-interaction phases such as common-envelope evolution, commonly parameterized by \(\alpha\) (see, e.g., \citealt{Santoliquido:2020axb,Broekgaarden:2021efa}). Plausible variations in these ingredients can shift the local normalization by factors of \(\sim 2\)--\(3\) (or more), even when calibrated to the same GW dataset. Taken together, these considerations place our adopted normalization within the range compatible with current observational and theoretical uncertainties.

Because we adopt a physical ab initio model, we do not rescale the merger-rate history to force the normalization of Eq.~\eqref{z02bbh} to match a specific local value, as such a rescaling would generally produce unphysical behavior at higher redshifts. Nevertheless, since current 2G detectors predominantly probe low redshifts (see Figure~\ref{fig:zplot}), we estimate that our predicted event counts for 2G networks would decrease by \(\sim 20\%\)--\(30\%\) if the merger-rate model were adjusted so that the normalization of Eq.~\eqref{z02bbh} lay within \([17.9,\,44]~\mathrm{Gpc}^{-3}\,\mathrm{yr}^{-1}\). We plan to update the data products associated with this work as new and improved merger-rate simulations become available.

The empirical normalization of the BHNS merger rate is currently subject to significant uncertainties, primarily due to the small sample of observed events. The observationally derived 90\%\ credible interval spans \([7.8,\,140]~\text{Gpc}^{-3}\,\text{yr}^{-1}\) \citep{KAGRA:2021duu}. Using the BHNS parameters listed in Table~\ref{tab:parameters}, our model predicts a local merger rate density at \( z=0.2 \) given by:
\begin{align} \label{z02bhns}
\int \mathcal{R}(z=0.2,M, \theta_{\rm BHNS})\,\mathrm{d}\log_{10}M = 20.3~\text{Gpc}^{-3}\,\text{yr}^{-1}\,,
\end{align}
comfortably within the observational constraints.

\subsubsection{BNS merger rate}\label{sec:bnsmerger}

For BNS mergers, the modeling approach is slightly different, primarily in the way the host galaxy mass dependence is incorporated. The dependence is summarized by a Gaussian-like function in logarithmic space to capture the variation across the wide range of host galaxy masses.
Thus, the total merger rate density \(\mathcal{R}\) for BNS mergers is expressed as:
\begin{align}
\label{SourceMRD_NS}
\mathcal{R}(z, M, \theta_{\rm NS}) &\!=\! \psi \!\times\! \exp \!\left\{ \! -0.5 \left[\frac{\log_{10}\left(\log_{10}\left(\frac{10^{ \nu(z)}}{M} \right) - L_c \right)}{\sigma_{\mathcal{_R}}} \right]^2 \right\} ,
\end{align}
where $\nu(z) = \frac{\nu_m}{1+Kz^2}$ displaces the limit of the distribution with redshift. The parameters governing the redshift evolution and the stellar mass distribution of the host galaxy are denoted by: 
\begin{align} \label{parNS}
\theta_{\rm NS} = \{ N_{\mathcal R}, A, B, C, \nu_m, K, L_c, \sigma_{\mathcal{_R}} \} ,
\end{align}
and their best-fit values are listed in Table~\ref{tab:parameters}.

Due to the limited number of observed BNS mergers (only two confirmed events), empirical constraints on their merger rate remain highly uncertain, with a wide 90\% credible interval of $[10, 1700] \text{ Gpc}^{-3} \text{ yr}^{-1}$ reported by \citet{KAGRA:2021duu}. Using the best-fit parameters from Table~\ref{tab:parameters}, we obtain:
\begin{align} \label{z02bns}
\int \mathcal{R}(z=0.2,M, \theta_{\rm NS}) , \mathrm{d}\log_{10}M = 888 \text{ Gpc}^{-3} \text{ yr}^{-1},
\end{align}
which comfortably lies within the broad empirical constraints.


Figure~\ref{Partial_MRD} illustrates the mass and redshift dependence of the best-fit merger rate densities for BBH, BHNS, and BNS systems. 
At fixed redshift (left panels), BBH and BHNS mergers display a relatively flat distribution at low stellar masses and a pronounced peak around 
$8 \times 10^{10}\,\mathrm{M}_\odot$ that becomes more prominent at low redshift. This feature reflects the cumulative effect of galaxy mergers occurring during the delay between BCO formation and coalescence, a period over which the host galaxy mass increases and its star formation rate and metallicity can evolve significantly. 
In contrast, BNS mergers exhibit a sharper peak at higher stellar masses, which shifts with redshift, indicating a distinct dependence on host-galaxy properties.
At fixed stellar mass (right panels), the merger rates peak around $z \approx 1$, broadly consistent with the cosmic star formation rate density, which reaches its maximum at $z \approx 2$ in the model of \citet{Madau:2014bja}. 
For a detailed discussion of these trends and their astrophysical interpretation, we refer the reader to \citet{Santoliquido:2022kyu}.

\subsubsection{Detector-Frame Merger Rate}

To convert the source-frame merger rate density into the detector frame rate, we use the following expression:
\begin{align} \label{det-ratez}
R(z,M, \theta_\text{cosmo}, \theta_{\rm BCO}) =  \frac{\mathcal R(z,M, \theta_{\rm BCO})}{(1+z)} \frac{\d V}{\d z} \,,
\end{align}
where $\theta_{\rm BCO}$ refers to either Eq.~\eqref{parBH} or \eqref{parNS}, and the redshift volume element \(\frac{\d V}{\d z}\) depends on cosmological parameters:
\begin{eqnarray} \label{volume}
 \frac{\d V}{\d z}= \frac{1}{E(z)}\left(\frac{c}{H_0}\right)^3\left[\int^{z_j}_0\frac{\d \bar z}{E(\bar z)}\right]^2 \Delta \Omega \,,
\end{eqnarray}
where the universe expansion, in the case of flat $\Lambda$CDM, is given by
\begin{align}
E(z) \equiv H(z)/H_0= \sqrt{\Omega_\text{m} (1+z)^3 + 1-\Omega_\text{m}} \,,
\end{align}
and $\Delta \Omega=\sky\text{ deg}^2$ corresponds to the sky area covered by the \texttt{CosmoDC2} catalog described in Section~\ref{Cosmo_DC2_sec}.
The cosmological parameters are denoted by:
\begin{align}
\theta_\text{cosmo} = \{ H_0, \Omega_\text{m}  \} \,.
\end{align}

\subsubsection{Number Counts}

\begin{figure}
\centering 
\includegraphics[trim={0 0 0 0}, clip, width= 1 \columnwidth]{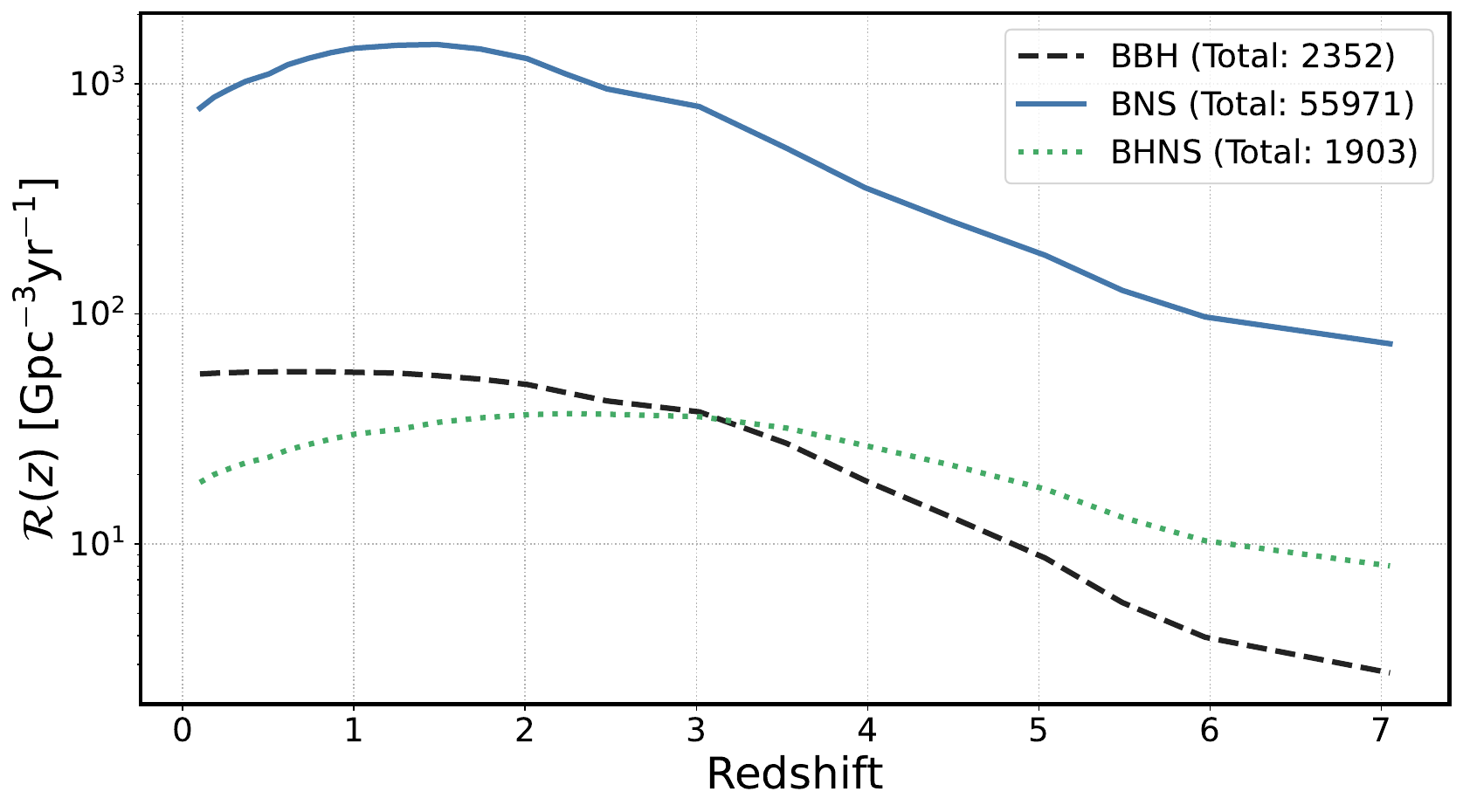}
\caption{Total source-frame merger rate as a function of redshift, obtained by integrating over the full mass range. The legend indicates the expected number of merger events within a sky area of 440 square degrees over a period of 10 years.}
\label{total_mrd}
\end{figure}

Using this formalism, we compute the expected number of merger events in a redshift bin \(\Delta z\) for host galaxies with stellar masses within a bin \(\Delta \log_{10}(M)\) over an observation period \(\Delta t\):
\begin{align} \label{Nij}
N_{ij} =  R(z_j, M_i, \theta_\text{cosmo}, \theta_{\rm BCO}) \, \Delta \log_{10}(M_i) \, \Delta z \, \Delta t \,.
\end{align}
In the following, we adopt \(\Delta t = 10\) yr.  
From this equation, one can compute the number of events per mass or redshift bin, as well as the total number of coalescences, given by
\begin{align}
N_\text{tot} = \sum_{i j } N_{ij} \,.
\end{align}
Figure~\ref{total_mrd} illustrates the source-frame merger rate as a function of redshift, obtained by integrating over the full mass range. The legend displays \(N_\text{tot}\).

\subsubsection{LSST Galaxy Catalog}
\label{Cosmo_DC2_sec}

We use the synthetic galaxy catalog \texttt{CosmoDC2},\footnote{\url{https://github.com/LSSTDESC/cosmodc2}} part of the LSST DESC's Data Challenge 2 \citep[DC2,][]{LSSTDarkEnergyScience:2019hkz}, which is produced by populating halos with galaxies through a combination of empirical and semi-analytic models. Specifically, we use the extra-galactic catalog version \texttt{CosmoDC2\_v1.1.4},\footnote{\href{https://portal.nersc.gov/project/lsst/cosmoDC2/_README.html}{\tt portal.nersc.gov/project/lsst/cosmoDC2/\_README.html}} which covers \sky deg$^2$ of the southern sky (see Table~\ref{tab:parameters_priors}), spans a redshift range extending up to $z \approx 3$ and is complete to a magnitude depth of 28 in the $r$-band. We access the \texttt{CosmoDC2} catalog via the \texttt{gcr-catalogs} 1.10.1 python package.\footnote{\url{https://github.com/LSSTDESC/gcr-catalogs/releases}}
This catalog is based on the Outer Rim simulation \citep{Heitmann:2019ytn}, which covers a volume of $(3 \text{ Gpc}/h)^3$ sampled with $10240^3$ tracer particles, leading to a mass resolution of $1.9 \times 10^9 M_\odot /h$. The fiducial cosmology is the flat $\Lambda$CDM model with $\Omega_{\rm cdm}=0.22$, $\Omega_{\rm b}=0.0448$, $n_s=0.963$, $h=0.71$, $\sigma_8=0.8$, which we will assume throughout this work.%
\footnote{The fiducial cosmology used by \citet{Santoliquido:2022kyu} differs slightly from the one adopted here; however, the resulting error is expected to be subdominant compared to the systematic uncertainties in modeling BCO mergers.}
The \texttt{CosmoDC2} catalog provides true values for key galaxy properties, including right ascension (RA), declination (DEC), redshift, stellar mass, morphology, spectral energy distributions, broadband magnitudes across multiple filters, host halo properties, and weak lensing shear, totaling over 500 attributes.

\subsubsection{Tiling}
\label{sec:tiles}

\begin{figure}
\centering 
\includegraphics[trim={0 1.6cm 0 5.5cm}, clip, width= 1 \columnwidth]{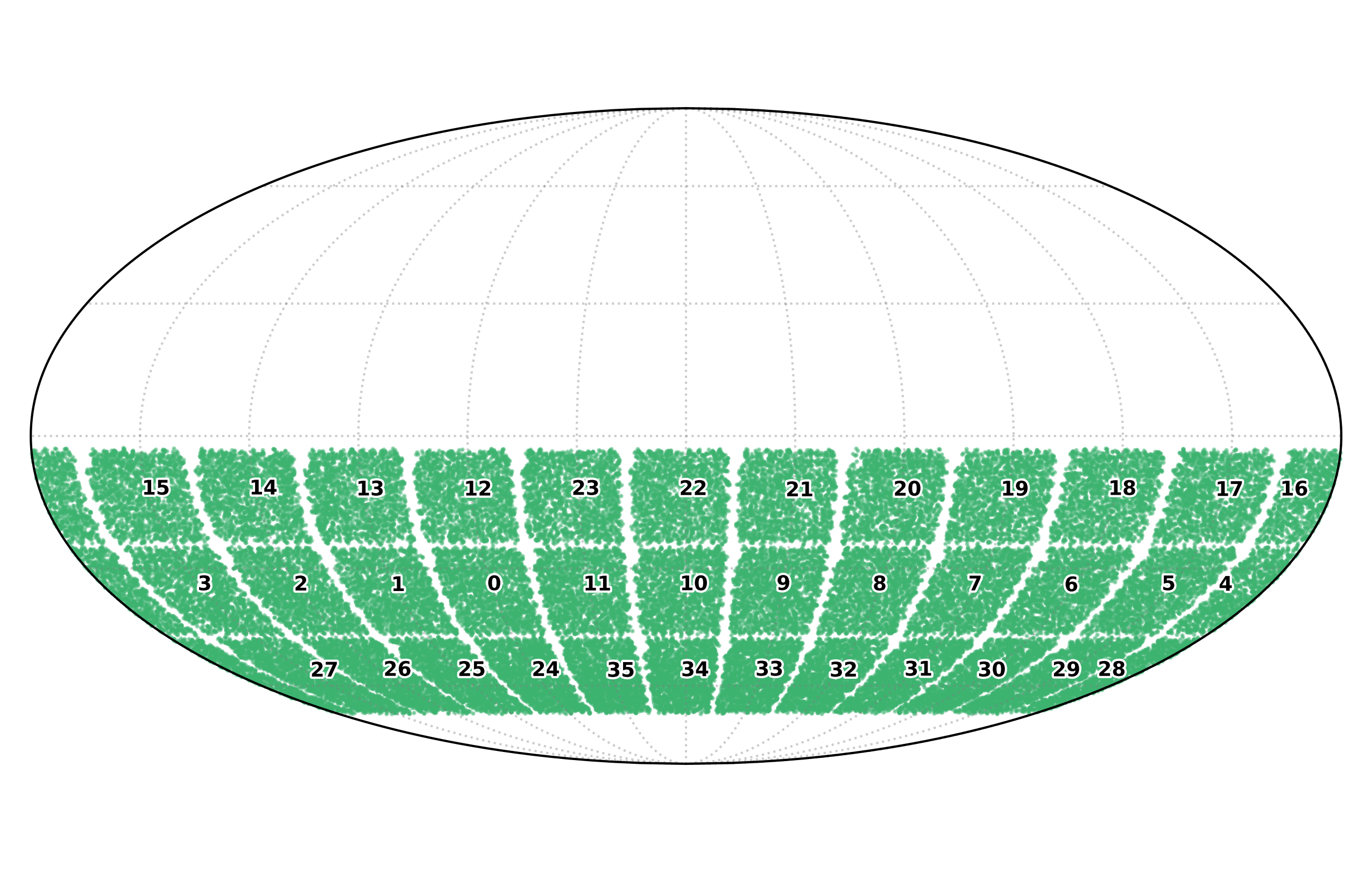}
\caption{Tiling scheme used to extend the \texttt{CosmoDC2} sky coverage to match the LSST survey area. The replicated tiles do not strictly follow the LSST footprint but are arranged to account for variations in the network pattern function across the sky while preserving a realistic sky distribution for dark siren cosmology studies.}
\label{fig:tiling}
\end{figure}

Since the sky coverage of \texttt{CosmoDC2} is limited to \sky deg$^2$, while LSST will observe approximately 18,000 deg$^2$ \citep{LSST:2008ijt}, we implement a tiling scheme to generate a catalog that represents the final LSST dataset. This approach enables forecasting of synergies relevant to dark siren cosmology.  

To achieve the required coverage, we replicate the original \texttt{CosmoDC2} tile 36 times in the southern hemisphere, arranging them in three rows of 12 tiles along right ascension, yielding a total area of 15,833 deg$^2$. Figure~\ref{fig:tiling} illustrates the adopted tiling configuration.

\begin{figure*}
\centering
\includegraphics[width=\textwidth]{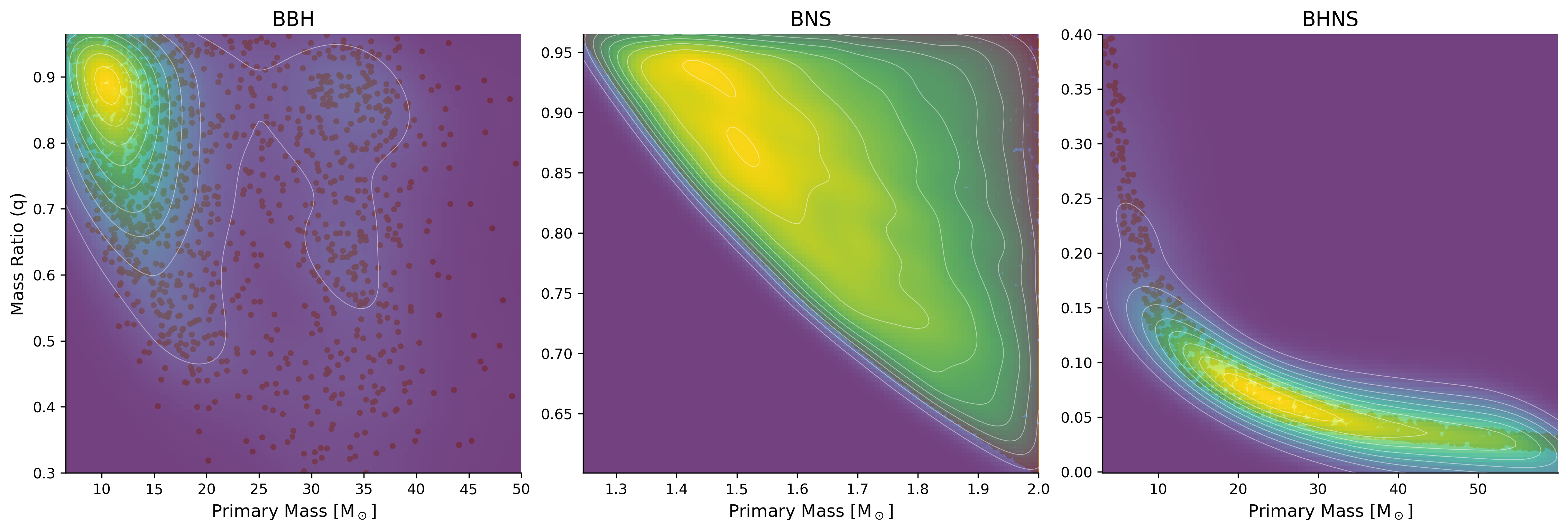}
\caption{Density distribution of primary mass \(m_1\) and mass ratio \(q\) for the simulated events. See Section~\ref{sec:mass-dist} for details.}
\label{fig:mass-dist}
\end{figure*}

This tiling approach does not aim to precisely match the LSST footprint, but rather to capture variations in the detector network pattern function across the sky, ensuring realistic sky coverage. Moreover, replicating the large-scale structure of \texttt{CosmoDC2} provides a reasonable approximation, since a significant fraction of mergers exhibit sky localization uncertainties \(\Delta \Omega \ll \sky \, \text{deg}^2\) (see Figure~\ref{DL-omega_err_net}), rendering the tiling impact negligible at the angular scales relevant for statistically associating a GW signal with potential LSST host galaxies.\footnote{Note that our approach is to include only events whose true host lies within the LSST footprint. Events originating outside the footprint may still partially overlap the survey boundary and thus be relevant for analyses sensitive to catalog incompleteness. In this first version of the catalog, we neglect such border effects.}
Note, however, that our catalog is not suitable for the large-scale structure approach of \citet{Ferri:2024amc}, which exploits correlations on large angular scales.

\subsubsection{Sampling Host Galaxies}

To sample host galaxies for BCO mergers, we follow the distributions described in Sections~\ref{sec:bbhmerger} and~\ref{sec:bnsmerger}. The galaxy catalog is partitioned into bins based on redshift (\(z_j\)) and stellar mass (\(M_i\)), with each bin containing thousands of galaxies. From these, we randomly select \(N_{ij}\) galaxies, as determined by Eq.~\eqref{Nij}, to serve as hosts for the simulated events.  
We assume a one-to-one correspondence between each selected host galaxy and a gravitational-wave event. This assumption, while simplifying the modeling process, is justified by the expectation that the rate of detectable mergers remains sufficiently low relative to the total galaxy population.
This procedure systematically defines the sky positions (right ascension and declination) and luminosity distances of simulated gravitational-wave events, assuming a specific cosmology. Over an observation period of 10 years, we estimate the total number of detectable mergers, as illustrated in Figure~\ref{total_mrd}.

\subsection{Binary Mass Distributions}
\label{sec:mass-dist}

The masses of compact objects play a crucial role in the simulation of mergers. Below, we outline the mass distributions adopted for each type of BCO considered in this study.

\subsubsection{Mass Distribution for BBH}

For the BBH mass distribution, we follow the modeling of \citet{Iacovelli:2022bbs}\footnote{A similar analysis can be found in \citet{KAGRA:2021duu}.}, which provides analytical fits based on the analysis of gravitational-wave events detected during the O1, O2, and O3 observing runs.\footnote{\url{https://zenodo.org/records/5655785}}
The compound distribution on the primary mass $m_1$ and  the mass ratio $q = m_2/m_1$ is given by
\begin{align} \label{BBH-m1q}
P(m_1, q| \theta_{m_1}, \theta_q) =&  P_{m_1}(m_1| \theta_{m_1})\times P_q(q|m_1, \theta_q) \,,
\end{align}
where the explicit expressions can be found in the Appendix~A2 of \citet{Iacovelli:2022bbs}.
Figure~\ref{fig:mass-dist} (left panel) illustrates the distribution of $m_1$ and $q$, highlighting their correlation. The presence of two local maxima is evident: a higher one at $\sim 5 M_\odot$ and a lower one at $\sim 33 M_\odot$.
This bimodal structure is consistent with the parameter-free analysis of \citet{Ray:2023upk,Callister:2023tgi}.

In this work, we assume a redshift-independent mass distribution. The non-parametric Bayesian analysis of \citet{Rinaldi:2023bbd} reports evidence for redshift evolution in both the primary-mass and mass-ratio distributions; however, this trend is not supported by the analysis of \citet{Heinzel:2024hva}.
If the redshift evolution suggested by \citet{Rinaldi:2023bbd} holds, BBH mergers would be typically more massive at higher redshift, increasing the detectability of high-redshift events and thus raising our forecasted detection counts at the highest redshifts.

\subsubsection{Mass Distribution for BNS and BHNS}
\label{NS-mass-dist}

The mass distributions for BNS and BHNS mergers exhibit notable differences compared to those of BBH, due to the limited number of observed events and differing astrophysical origins. These distributions are essential for accurately modeling the population characteristics and expected GW signals from these sources.

The range of masses for BNS is relatively narrow. Observed Galactic binary neutron stars have individual masses typically between 1 and 2 \(M_{\odot}\), often well described by a narrow Gaussian centered around 1.33 \(M_{\odot}\) \citep{Farrow:2019xnc}. However, gravitational wave observations to date suggest that neutron star masses in binary mergers do not follow a single-peaked distribution. Instead, the inferred mass distribution for neutron stars observed in gravitational waves is broader \citep{KAGRA:2021duu}, with more support for high-mass neutron stars compared to the Galactic NS population. Guided by gravitational wave inferences, the mass range for BNS covers values between $m_{\rm min} = 1.2 M_{\odot}$ and $m_{\rm max} = 2 M_{\odot}$, following a power-law distribution with smoothed edges:
\begin{align}
\label{M_NS}
P_{m_{\rm NS}} (m | \alpha_n) &\propto S_{\rm NS}(m)  \cdot m^{\alpha_{N}}  \,, \\
S_{\rm NS}(m) &= \frac{1}{1 + e^{-b(m - m_{\text{min}})}} \cdot \frac{1}{1 + e^{b(m - m_{\text{max}})}} \,,
\end{align}
where \( S_{\rm NS} \) is a smoothing function that exponentially damps the distribution at the edges of the mass range. We have assumed a power law with a declining slope of $\alpha_N = -2.1$ \citep{KAGRA:2021duu} and a smoothing coefficient of $b = 15$.

For black holes accompanying neutron stars, given the limited available data, we assume a uniform distribution spanning from \(3\) to \(60\) \(M_{\odot}\):
\begin{align}
\label{M_BH_with_NS}
P_{m_{\rm BH}} (m) &= U(3 M_{\odot}, 60 M_{\odot}) \,,
\end{align}
so that the full mass distribution for these binaries is:
\begin{align}
P(m_1, m_2) \!=\!\!
\begin{cases}
\! P_{m_{\rm NS}} (m_1 | \alpha_n) \, P_{m_{\rm NS}} (m_2 < m_1 | \alpha_n), & \text{if BNS} \\
\! P_{m_{\rm BH}} (m_1) \, P_{m_{\rm NS}} (m_2 | \alpha_n). & \text{if BHNS}
\end{cases}
\end{align}
Figure~\ref{fig:mass-dist} illustrates these distribution.

\subsection{Spin Distributions}
\label{spins_sec}

The spin of each compact object is fully characterized by its magnitude (\(\chi\)) and direction (\(\theta, \phi\)), which are modeled as follows.

\subsubsection{Magnitude}

The dimensionless spin parameter is defined as  
\begin{align}
\chi = \frac{c J}{G M^2} \,,
\end{align}
where \( J \) is the spin angular momentum of the compact object. For a Kerr black hole, the spin is bounded by  $\chi \leq 1$.

In a BBH system, for each black hole, we sample the spins using the distributions presented in \citet{LIGOScientific:2018jsj}. The spin magnitudes, \(\chi_1\) and \(\chi_2\), follow a Beta distribution, defined over the interval \([0, 1]\):
\begin{equation}
\label{spin_BH}
P_{\chi_{\rm BH}}(\chi | \alpha, \beta) = \frac{\chi^{\alpha - 1}(1 - \chi)^{\beta - 1}}{B(\alpha, \beta)} \,,
\end{equation}
where \(\alpha\) and \(\beta\) are the shape parameters, and \(B(\alpha, \beta)\) is the Beta function. 
We adopt \(\alpha=1.6\) and \(\beta=4.12\), following \citet{Iacovelli:2022bbs}.

Observations by \citet{Burgay:2003jm} indicate that although some known neutron star systems exhibit relatively high dimensionless spins (\(\chi \lesssim 0.5\)), the majority have spins below \(0.05\) \citep{Zhu:2017znk}. Motivated by this, we adopt a distribution that favors slowly rotating neutron stars, sampling spins from a truncated Gaussian centered at zero (\(\mu_\chi = 0\)) with standard deviation \(\sigma_\chi = 0.1\):
\begin{equation}
\label{spin_NS}
P_{\chi_{\rm NS}}(\chi | \mu_\chi, \sigma_\chi) = 2 \, \mathcal{G}(\chi | \mu_\chi, \sigma_\chi) \,.
\end{equation}
This choice provides a conservative and robust prior, allowing for moderate spin values without strongly constraining the inference or introducing significant degeneracies. As shown in \citet{Abbott:2018exr}, adopting broader spin priors does not bias parameter recovery but rather broadens posterior distributions.

\subsubsection{Direction}

Following \citet{LIGOScientific:2018jsj}, the angle \(\theta\) follows a mixture of two distributions: an isotropic component and a component favoring alignment with the \(z\)-direction, assuming the \(z\)-axis aligns with the binary's angular momentum. The cosines of the spin angles, \(\cos(\theta_1)\) and \(\cos(\theta_2)\), are sampled from the joint probability density function:
\begin{equation}\label{spin_dir}
P(\cos \theta_1, \cos \theta_2 | \zeta, \sigma) \!=\! \frac{1 \!-\! \zeta}{4} \!+\! \frac{2\zeta}{\pi} \frac{\exp\!\! \left[\! {-\frac{(1 - \cos\theta_1)^2}{2\sigma^2} \!-\! \frac{(1 - \cos\theta_2)^2}{2\sigma^2}}\right ]}{\sigma^2 \operatorname{erf}^2(\sqrt{2}/\sigma) } ,
\end{equation}
where \(\zeta\) controls the relative weight of the two components, and \(\sigma\) defines the width of the Gaussian distribution.  
We adopt \(\zeta = 0.66\) and \(\sigma=1.5\), following \citet{Iacovelli:2022bbs}.\footnote{Note, however, that \citet{Iacovelli:2022bbs} restrict their analysis to the spin-aligned case.}

The azimuthal spin angles \(\phi_1\) and \(\phi_2\) are uniformly distributed over the interval \([0, 2\pi]\):
\begin{equation}
P(\phi) = \frac{1}{2\pi}\,.
\end{equation}

\subsection{Further GW Parameters}
\label{Further_para_sec}

Additional parameters are required to fully characterize a gravitational wave signal from a merger. These include the Euler angles that define the source orientation relative to the observer: the inclination angle \(\iota\), the polarization angle \(\psi\), and the coalescence phase \(\Phi_c\). All of these parameters are typically sampled from uniform distributions, following a standard approach as in \citet{Iacovelli:2022bbs}.


Another important parameter is the coalescence time, $t_c$, defined relative to a reference detector, which primarily affects the waveform as a phase shift. The corresponding coalescence times at other detectors depend on their geographical positions on Earth.  
In the main analysis, we set $t_c = 0$, as its effect is degenerate with changes in right ascension due to Earth's rotation. To assess the robustness of this assumption, given the non-uniform sensitivity of the detector network across the sky, we repeated the analysis with a random offset added to the right ascension of each event, uniformly distributed between $0^\circ$ and $360^\circ$. This procedure effectively simulates varying coalescence times. The results were consistent with the original analysis, confirming that our approximation is justified.

\begin{table}
\centering
\caption{Summary of the sampling distributions adopted in this work. Note that $m_1$ ranges are given in $M_{\odot}$.\label{tab:parameters_priors}}
\setlength{\tabcolsep}{5pt}
\renewcommand{\arraystretch}{1.6}
\begin{tabular}{|c|c|c|c|c|}
\hline
\textbf{Par.} & \textbf{BBH} & \textbf{BHNS} & \textbf{BNS} & \textbf{Range} \\ \hline
$m_1$ & Power Law & $ U(3 , 60)^{\eqref{M_BH_with_NS}}$ & $P_{m_{\rm NS}}^{\eqref{M_NS}}$ & [3/3/1, 100/60/2]  \\ \cline{1-1} \cline{3-3} \cline{4-4} \cline{5-5}
$q$, $m_2$ & + Peak$^{\eqref{BBH-m1q}}$ & $P_{m_{\rm NS}}^{\eqref{M_NS}}$ & $P_{m_{\rm NS}}^{\eqref{M_NS}}$ & [0, 1], [1, 2] $M_\odot$ \\ \hline
$z$ & $\mathcal{R}^{\eqref{SourceMRD}}$ & $\mathcal{R}^{\eqref{SourceMRD}}$ & $\mathcal{R}^{\eqref{SourceMRD_NS}}$ & [0, 3] \\ \cline{1-1} \cline{5-5}
$\phi$ & + & + & + & [49.5$^\circ$, 74.5$^\circ$] \\ \cline{1-1} \cline{5-5}
$\delta$ & \texttt{CosmoDC2} & \texttt{CosmoDC2} & \texttt{CosmoDC2} & [-45$^\circ$, -26$^\circ$] \\ \hline
$\iota$ & \multicolumn{3}{|c|}{Uniform in $\cos{\iota}$}  & [-1, 1] \\ \hline
$\psi$ & \multicolumn{3}{|c|}{Uniform} & [0, $\pi$] \\ \hline
$\phi_{c}$ & \multicolumn{3}{|c|}{Uniform} & [0, 2$\pi$] \\ \hline
$t_{c}$ & \multicolumn{3}{|c|}{Fixed} & 0 \\ \hline
$\chi_{1}, \chi_2$ & $P_{\chi_{\rm BH}}^{\eqref{spin_BH}}$ & $P_{\chi_{\rm BH}}^{\eqref{spin_BH}}$,  $P_{\chi_{\rm NS}}^{\eqref{spin_NS}}$ & $P_{\chi_{\rm NS}}^{\eqref{spin_NS}}$ & [0, 1] \\ \hline
$\theta_{1}, \theta_2$ & \multicolumn{3}{|c|}{$P(\cos \theta_1,\cos \theta_2)^{\eqref{spin_dir}}$}  & [0, $\pi$] \\ \hline
$\phi_{1}, \phi_2$ & \multicolumn{3}{|c|}{Uniform} & [0, 2$\pi$] \\ \hline
\end{tabular}
\end{table}

\subsection{Input catalog}

A BCO merger originating from a circular orbit is fully characterized by a set of 15 parameters, which describe both the intrinsic properties of the binary system and the extrinsic parameters related to its orientation and position with respect to the observer. We denote the full set of gravitational-wave source parameters as $\theta_{\rm gw}$, which can be separated into intrinsic and extrinsic subsets as follows:
\begin{align}
\theta_{\rm gw} &= \theta_{\rm gw}^{\rm int} \cup  \theta_{\rm gw}^{\rm ext}  \, , \label{15par}\\
\theta_{\rm gw}^{\rm int} & =  \left \{ \mathcal{M}_c^z, \eta, \chi_1, \theta_1, \phi_1, \chi_2, \theta_2, \phi_2 \right \} \,,  \label{15int}\\
\theta_{\rm gw}^{\rm ext}&= \left \{ D_L, \phi, \delta, \iota, \psi, \Phi_c, t_c \right \} \label{15ext}  \,, 
\end{align}
where \(D_L\) is the luminosity distance in Gpc, and the redshifted chirp mass \(\mathcal{M}_c^z\) and symmetric mass ratio \(\eta\) are defined as:
\begin{align}
\mathcal{M}_c &= \frac{(m_1 m_2)^{3/5}}{(m_1 + m_2)^{1/5}} \,,\\
\mathcal{M}_c^z &= (1+z) \, \mathcal{M}_c \,, \label{chirpmassz}\\
\eta &= \frac{m_1 m_2}{(m_1 + m_2)^2} \,.
\end{align}
We adopt \(\mathcal{M}_c^z\) and \(\eta\) as independent parameters since observations directly constrain these combinations of \(m_1\) and \(m_2\), and the posterior distribution is more Gaussian with respect to these parameters.
For this study, we do not include tidal deformability effects in neutron star mergers. Future iterations will incorporate this additional parameter to refine waveform modeling and enhance the astrophysical interpretation of detected events.

Table~\ref{tab:parameters_priors} summarizes the prior distributions used to sample the 15 parameters for generating synthetic GW catalogs. The sampled events collectively form the \emph{Input Catalog}, an essential component of the workflow illustrated in Figure~\ref{fig:flowchart}. The Input Catalog consists of 36 replicated sky tiles, following the methodology described in Section~\ref{sec:tiles}, with its detailed structure outlined in Table~\ref{tab:data-prod}.

\subsection{Gravitational-Wave Detector Networks}
\label{networks}

\begin{table}
\centering
\caption{Detector configurations. \label{tab:detectors}}
\setlength{\tabcolsep}{7pt}
\renewcommand{\arraystretch}{1.3}
\begin{tabular}{|c|c|c|c|c|c|}
\hline
\textbf{Detector} & \textbf{Arms} & \textbf{Lat.} & \textbf{Long.} & \textbf{X Arm} & \textbf{Y Arm} \\
\hline
LIGO Han.  & 4 km & 46.45\textdegree & -119.40\textdegree &  N36\textdegree W &  W36\textdegree S \\
\hline
LIGO Liv.  & 4 km & 30.56\textdegree & -90.77\textdegree &  W18\textdegree S &  S18\textdegree E \\
\hline
Virgo & 3 km & 43.51\textdegree & 10.43\textdegree &  N19\textdegree E &  W19\textdegree N \\
\hline
Kagra & 3 km & 36.41\textdegree & 137.31\textdegree &  E28.3\textdegree N &  N28.3\textdegree W \\
\hline
$\text{ET}_{\Delta}$ & 10 km & 50\textdegree & 6\textdegree &  $\triangle$ &   $\triangle$ \\
\hline
$\text{ET}_{L1}$ & 10 km & 50\textdegree & 6\textdegree & E30\textdegree N  &   N30\textdegree W  \\
\hline
$\text{ET}_{L2}$ & 10 km & 40.44\textdegree & 9.44\textdegree &  S0\textdegree E  &   E90\textdegree N  \\
\hline
$\text{CE}_1$ & 40 km & 38\textdegree & -78\textdegree &  S45\textdegree E &   E45\textdegree N \\
\hline
$\text{CE}_2$ & 40 km & -32\textdegree & 116\textdegree &  S0\textdegree E &   E0\textdegree N \\
\hline
\end{tabular}
\end{table}

\begin{figure}
\centering 
\includegraphics[trim={0 0 0 0}, clip, width= 1 \columnwidth]{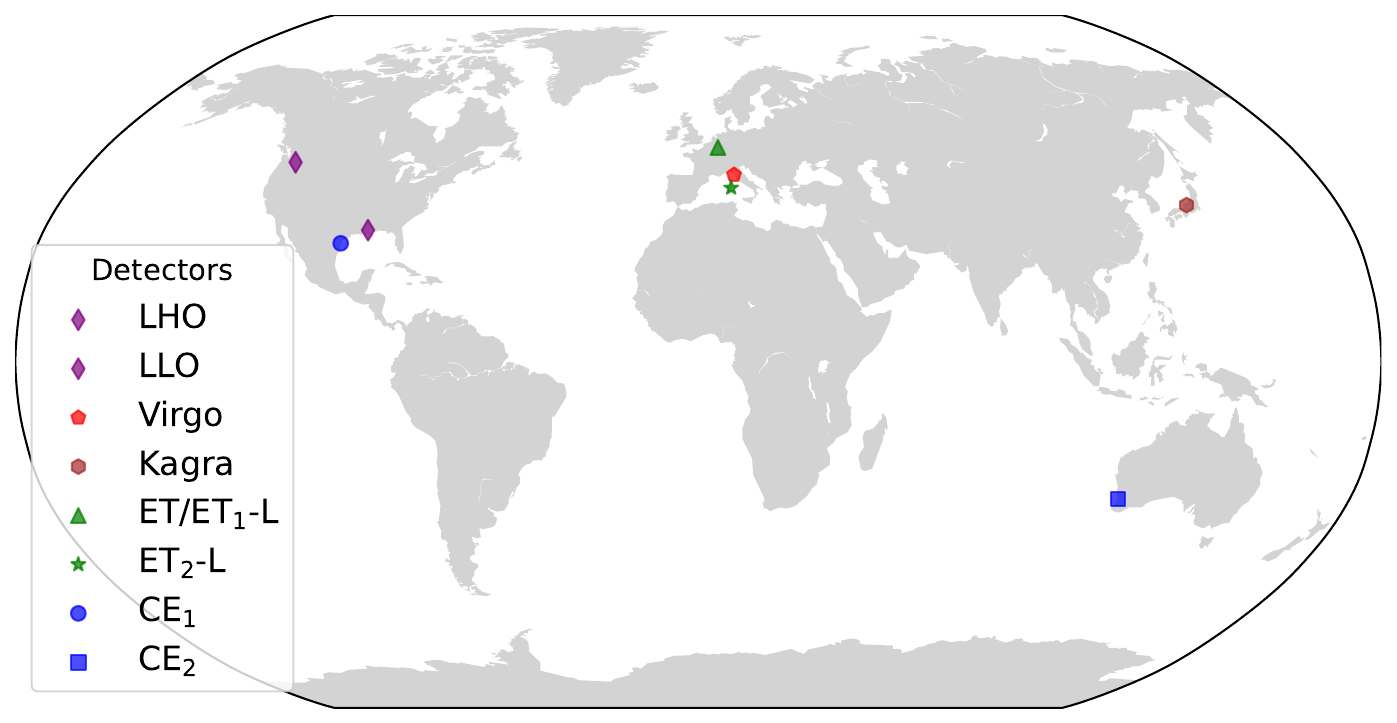}
\caption{Locations considered in this work for each detector.}
\label{location}
\end{figure}

\begin{figure}
\centering 
\includegraphics[trim={0 0 0 0}, clip, width= 1 \columnwidth]{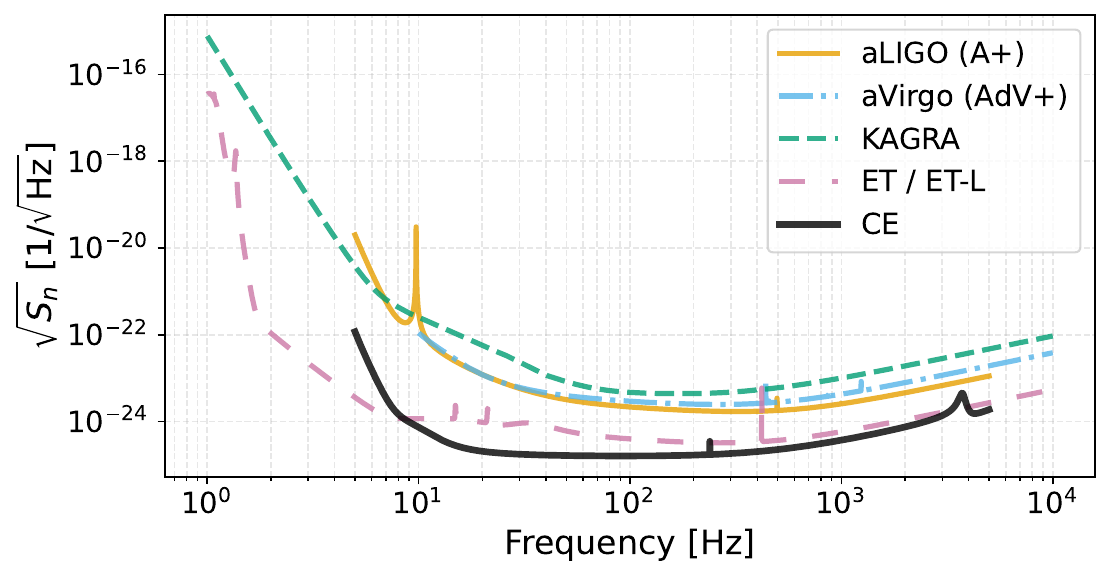}
\caption{Predicted sensitivity curves.}
\label{SensCurves}
\end{figure}

We consider multiple configurations of gravitational-wave detector networks. Below, we describe the second- and third-generation detectors included in this study.

GEO600 and LIGO-India are not included in our analysis. GEO600, due to its significantly lower sensitivity relative to current-generation detectors, does not contribute meaningfully to detection capabilities. LIGO-India, while a promising addition to the network, is expected to reach sensitivities comparable to LIGO and Virgo only in the coming years. Since our focus is on evaluating the impact of third-generation detectors on the current network, including LIGO-India in mixed configurations would not provide substantial improvements, while incurring additional computational cost. Nevertheless, LIGO-India may be included in future work as a complementary component.

\subsubsection{LIGO-Virgo-KAGRA Network}

The second-generation (2G) LIGO-Virgo-KAGRA (LVK) network is currently operational, and we consider its final design sensitivity.  
Figure~\ref{location} shows the locations of the three detectors, while Table~\ref{tab:detectors} provides their specifications.\footnote{\url{http://35.167.8.93/scientists/GRB051103/GRB051103-geometry.php}}
For this study, we use the sensitivity curves corresponding to the upcoming O5 observing run \citep{ETGW2024,KAGRA:2013rdx},\footnote{\url{https://dcc.ligo.org/LIGO-T2000012-v1/public}} expected to operate in the coming years, as shown in Figure~\ref{SensCurves}.

\subsubsection{Einstein Telescope}

The Einstein Telescope is a 3G gravitational-wave observatory designed to achieve significantly higher sensitivity using a 10 km interferometer\footnote{\url{https://apps.et-gw.eu/tds/ql/?c=14313}} \citep{ET:2019dnz}, see Figure~\ref{SensCurves}. Its construction has already secured significant funding, with the final site decision expected by 2027 and construction projected to begin shortly thereafter. The candidate sites currently under consideration are island of Sardinia, Italy, and the Euregio Meuse-Rhine region, located at the border of Belgium, Germany, the Netherlands, and the German state of Saxony\footnote{\href{https://physicsworld.com/a/physicists-set-to-decide-location-for-next-generation-einstein-telescope/}{Physics World (10 Sep 2025)}} \citep{ET-Site}.  

The ET will consist of three interferometers arranged in an equilateral triangle, with two detectors at each vertex. This configuration enables precise measurement of gravitational-wave polarization while ensuring sensitivity to all wave orientations, eliminating potential blind spots.

However, simulations indicate that the 60\textdegree~arm angle reduces the sensitivity of each interferometer, leading to the consideration of an alternative ET configuration. This alternative consists of two detectors with the same individual sensitivity and dimensions as the standard ET but arranged in an L-shaped configuration (90\textdegree~arm angle) and located in Sardinia and the Euregio Meuse-Rhine region. See Table~\ref{tab:detectors} for details\footnote{\url{https://apps.et-gw.eu/tds/?r=15601}} and Figure~\ref{location} for the locations.
In the following, we explore the potential impact of the alternative 2L setup \citep{Branchesi:2023mws,Maggiore:2024cwf}.

\subsubsection{Cosmic Explorer}

The Cosmic Explorer  is a proposed but not yet funded 3G ground-based gravitational-wave observatory \citep{Evans:2021gyd}, designed to detect the majority of gravitational-wave signals from binary compact object mergers. It features an L-shaped design similar to that of the LIGO detectors but with arms that are ten times longer (see Table~\ref{tab:detectors}, \citealt{Evans:2021gyd}).
The increased arm length enhances the signal amplitude while maintaining a low noise level,\footnote{\url{https://dcc.ligo.org/LIGO-P1600143/public}} see Figure~\ref{SensCurves}.

The exact locations of the Cosmic Explorer observatories have not yet been finalized; the provisional sites are near the current Hanford and Livingston LIGO facilities.\footnote{\url{https://cosmicexplorer.org/celocations.html}} 
\citet{Gupta:2023lga,deSouza:2023gjv} indicate that the best constraints on luminosity distance and sky localization are obtained when detectors are either co-located or placed at nearly antipodal positions, due to the increased baseline for triangulation.
Accordingly, for our analysis we adopt the provisional configuration shown in Figure~\ref{location}, placing CE$_1$ near Livingston and CE$_2$ in Western Australia, near Perth. This setup optimizes the network's parameter estimation capabilities while ensuring at least one 3G detector in the Southern Hemisphere.
In the following, when a network with a single CE is considered, it refers to the one located in the United States.

\subsubsection{Duty Cycle}\label{sec:dc}

A detector network operating continuously, producing useful data at all times, would have a duty cycle (DC) of \(100\%\). However, this is not realistic, and it is essential to account for the effects of a non-ideal DC. While catalogs assuming a \(100\%\) DC are unrealistic, they remain useful for isolating the impact of waveform models and detector networks without introducing observational biases related to the DC.

We simulate a non-ideal DC using a Monte Carlo approach. For instance, in the case of a \(70\%\) DC, each detector in the network is assigned a random number between 0 and 1 for every gravitational-wave event. Only detectors with a random number below the adopted DC are considered operational for that event.

\subsection{Waveform Models}
\label{wave-models}

The development of physically rich, accurate, and fast inspiral–merger–ringdown models capable of capturing the full details of a BCO coalescence is an active area of research.  
A Post-Newtonian (PN) expansion is typically combined with Numerical Relativity (NR) simulations to achieve greater accuracy.  
When comparing models with real data, completeness and accuracy are essential to avoid biases in the inference of coalescence parameters. In forecasting, improved modeling allows for sensitivity to richer physical phenomena compared to analyses using simpler waveform models, generally leading to stronger constraints on the parameters.  
Below, we summarize the waveform models (approximants) employed in this study:
\begin{itemize}


\item \texttt{IMRPhenomXAS} is a full inspiral–merger–ringdown model calibrated with NR simulations, assuming non-precessing spins.

\item \texttt{IMRPhenomXHM} extends the standard quadrupole signal by including subdominant multipoles, improving the accuracy of the multipole expansion. While this refinement introduces a slight increase in computational cost compared to simpler models like \texttt{IMRPhenomXAS}, it significantly enhances the model’s precision. In this study, we examine the impact of these higher modes on BBH merger detections.

\item \texttt{IMRPhenomXP} incorporates the effects of spin precession due to misaligned spins using PN expansions \citep{Pratten:2020ceb}. The additional computational cost is relatively small compared to \texttt{IMRPhenomXAS}.

\item \texttt{IMRPhenomXPHM} is the most comprehensive waveform model considered in this study, as it accounts for both spin precession and higher-order multipoles.

\end{itemize}


The waveforms listed were developed for BBH systems. The primary difference between BBH waveforms and those adapted for BNS or BHNS systems is the inclusion of tidal deformability parameters \( \Lambda_i \), which modify the waveform phase through terms suppressed by factors proportional to \( q^4 \) \citep{Vines:2011ud}. Consequently, tidal effects are negligible for BHNS systems but can impact the waveform of BNS systems  (see Figure~\ref{fig:mass-dist}).

Accurate treatment of tidal effects requires detailed knowledge of the neutron star equation of state, which is not yet incorporated into the current analysis. Moreover, the waveform generation and inference code \texttt{GWDALI}, employed throughout this work, does not currently support tidal corrections. For these reasons, we use the \texttt{IMRPhenomXHM} waveform for BNS systems and \texttt{IMRPhenomXPHM} for BHNS systems, while deferring the inclusion of tidal effects to future updates.

\texttt{IMRPhenomXHM} is a suitable choice for BNS systems, offering reduced computational cost while incorporating higher-order modes in the quadrupole expansion. This enhances parameter estimation and helps mitigate degeneracies. Spin precession is not significant for BNS systems, as neutron stars in binaries typically exhibit low spin magnitudes. For BHNS systems, \texttt{IMRPhenomXPHM} is more appropriate, as it includes both higher-order modes and spin precession effects, which are generally more relevant due to the typically higher spins of black holes and larger mass ratios.

\subsection{Signal-to-Noise Ratio}
\label{sec:snr}

The signal-to-noise ratio (SNR) quantifies the detectability of a gravitational-wave signal. Assuming  \( S_n(f) \), the SNR \( \rho \) for a given signal \( d(t) \) is defined as \citep{maggiore2008gravitational}:
\begin{align} \label{SNR}
\rho = \langle d, d \rangle^{1/2} \,,
\end{align}
where the inner product \( \langle a, b \rangle \) is given by:
\begin{align}
\langle a, b \rangle = 4 \text{Re} \int_0^\infty \frac{\tilde{a}^*(f) \tilde{b}(f)}{S_n(f)} \, \d f \,.
\end{align}
Here, \(\tilde{a}(f)\) and \(\tilde{b}(f)\) denote the Fourier transforms of the time-domain signals \( a(t) \) and \( b(t) \), respectively, and \(\text{Re}\) denotes the real part of the complex integral.

For a network of detectors indexed by \( i \), the network SNR is obtained by summing the individual detector SNRs in quadrature:
\begin{align}\label{rhonet}
\rho_{\rm net} = \sqrt{\sum_i \rho_i^2} \,.
\end{align}
Following \citet{Chen:2014yla, KAGRA:2021vkt}, we define an event as detectable if its network SNR exceeds the threshold:
\begin{align}\label{rhocut}
\rho_{\rm net} \geq \rho_{\rm cut} = 12 \,.
\end{align}


\subsubsection{Network Response}

A gravitational-wave detector measures a linear combination of the two independent polarization states of the wave, $d_+$ and $d_\times$, weighted by the pattern functions \( F_+ \) and \( F_\times \), which encode the detector's response to different sky locations. The measured strain is given by  
\begin{equation}
d(t) = F_+(\phi, \delta, \psi) d_+(t) + F_\times(\phi, \delta, \psi) d_\times(t),
\end{equation}
where \( \phi \) and \( \delta \) are the right ascension and declination of the source, and \( \psi \) is the polarization angle. These pattern functions depend on the detector's orientation and position on Earth and vary across the sky, making a network of detectors necessary to fully reconstruct the gravitational-wave signal.

In order to check for blind or less sensitive sky directions it is useful to plot the quadratic sum of pattern functions for the network under examination.
As the pattern functions of a detectors only depend on the geometrical configuration of the interferometer, when combining more detectors it is meaningful to weight the combination via the square of the SNR of Eq.~\eqref{SNR} for a flat signal:
\begin{align}
w =  4  \int_0^\infty \frac{1}{S_n(f)} \, \d f \,,
\end{align}
so that the effective quadratic sum of the pattern functions for the network is:
\begin{align}\label{Fnet}
F_{\rm net} =  \sqrt{\sum_i w_i \left( F_{+,i}^2 + F_{\times,i}^2 \right) } \,.
\end{align}

\subsection{Statistical Inference}
\label{inference}

Bayesian inference is performed using the following likelihood function:
\begin{align}
\mathcal{L}(d | h) \propto \exp\left(-\frac{1}{2} \langle d-h, d-h \rangle\right) \,, \label{likefull}
\end{align}
where \( h(t) \) is the theoretical waveform model intended to describe the observed data \( d \).  
When multiple detectors are involved, their likelihoods are combined multiplicatively.

The template \( h(t) \) that maximizes the likelihood is considered the best match to the data, assuming Gaussian noise and an accurately known spectral noise density \( S_n(f) \).  
Simulating a single merger incurs significant computational cost, particularly when accounting for all 15 free parameters in Eq.~\eqref{15par}.

\subsubsection{Fisher Approximation}
\label{fisher}

Exploring the full parameter space of Eq.~\eqref{likefull} is computationally expensive. To mitigate this, we apply an approximation using the Fisher matrix formalism, which significantly reduces computational cost:
\begin{align}
\mathcal{F}_{ab} = 4  \text{Re} \int^{\infty}_0 \frac{\partial_a \tilde{h}^{*}(f) \, \partial_b \tilde{h}(f)}{S_n(f)} \, \d f \,,
\end{align}
where \( \mathcal{F}_{ab} \) denotes the Fisher matrix elements, and the indices \( a, b \) run over the gravitational wave parameters in Eq.~\eqref{15par}.  
Once the Fisher matrix is computed, the covariance matrix of the gravitational wave parameters is obtained by inverting it:
\begin{align}
\mathcal{C} =\mathcal{F}^{-1} \,.
\end{align}

We employ the code \texttt{GWDALI} v0.1.4 \footnote{\url{https://github.com/jmsdsouzaPhD/GWDALI}} \citep{deSouza:2023ozp} for waveform generation and inference. \texttt{GWDALI} supports the DALI (higher-order Fisher) approximation to the full posterior \citep{Sellentin:2014zta}. However, we limit the analysis to the standard Fisher approximation, as the implementation of precise higher-order derivatives required for DALI is not yet complete.

Despite its computational efficiency, the Fisher approximation is not always reliable for all events. In some cases, numerical inversion of the Fisher matrix is imprecise, leading to a covariance matrix that does not correctly correspond to the inverse of the Fisher matrix \citep{deSouza:2023ozp}.  
To address this issue, we provide quality metrics for each detected event in the output catalogs to assess the reliability of the Fisher matrix inversion. Specifically, we evaluate the deviation of the product between the Fisher matrix and its numerical inverse from the identity matrix:
\begin{align}
\label{inverse_metric}
\max_{i,j} \left| (\mathcal{F} \mathcal{F}^{-1} - I)_{ij} \right| \,.
\end{align}
We consider the inversion reliable when this deviation is smaller than \( 10^{-3} \).

\subsubsection{MCMC}
\label{mcmc}

We perform multiple analyses, varying the network configuration, approximant, and duty cycle, with each analysis consisting of thousands of events. Given the computational cost, a full MCMC exploration is not feasible for all cases.  
Using the \texttt{emcee} sampler within \texttt{bilby} \citep{2019ApJS..241...27A,2020MNRAS.499.3295R}, analyzing a single synthetic event requires more than 10 hours on a modern computational node.  

Nonetheless, to assess the accuracy of the Fisher approximation, we conducted a full MCMC exploration for all BBH mergers in the ET+LVK network configuration. 
The results, presented in Appendix~\ref{ap:validation}, indicate that the Fisher approximation provides reasonable predictions. However, a full MCMC analysis is recommended for obtaining precise parameter estimates.
See \citet{Dupletsa:2024gfl} for a study evaluating the accuracy of the Fisher approach on actual LVK events. It is important to emphasize that the Fisher approximation remains particularly well-suited for assessing relative changes across different network configurations.

\subsubsection{Sky area}

In gravitational wave catalogs, the sky location of a source is specified by right ascension and declination. However, to optimize our analysis of sky localization, it is advantageous to use the angular area constrained in the sky for each event source. The likely region in the sky for an event’s location is defined by the area of an ellipse determined by the uncertainties in right ascension and declination, as well as their correlation.

As preented in \citet{Wen:2010cr}, the probability that an event falls outside the solid angle \(\Delta \Omega\) due to statistical error follows an exponential decay:
\begin{equation}
P(\Delta \Omega > x) \propto \exp\left(-\frac{x}{\Delta\Omega_s}\right),
\end{equation}
where the characteristic angular uncertainty \(\Delta\Omega_s\) is given by:
\begin{equation}
\Delta\Omega_s = 2\pi | \sin \delta| \sqrt{ \det \mathcal{C}_{\delta \phi} }\,.
\end{equation}
Here, \(\mathcal{C}_{\delta \phi}\) is the covariance matrix marginalized over all parameters except \(\delta\) (declination) and \(\phi\) (right ascension). The factor \( |\sin \delta| \) accounts for the sky projection effect in spherical coordinates.
For a confidence level of \(X\%\), the associated solid angle is:
\begin{equation}
\Delta \Omega_X = - \Delta \Omega_s \ln\left(1 - \frac{X}{100}\right) \,.
\end{equation}

\subsection{Output Catalog}

At this stage of the workflow in Figure~\ref{fig:flowchart}, we construct the \emph{Output Catalog}, which is determined by the selected detector network configuration (including duty cycle), waveform model, and inference method. 
The complete structure of the \emph{Output Catalog} is detailed in Table~\ref{tab:data-prod}.

\subsection{Kilonova Luminosity}
\label{sec:kn}

Neutron star mergers are of particular interest in cosmology, as they can act as bright sirens when their electromagnetic (EM) counterparts are detected alongside their gravitational-wave signals. These multimessenger observations provide an independent method for constraining cosmological parameters. 

A key challenge in modeling kilonova luminosity is the uncertainty in the neutron star equation of state (EOS), which influences the mass ejection dynamics and the resulting EM emission. Here, we outline the methodology used to estimate the luminosity of these EM counterparts.

\subsubsection{BHNS}
\label{sec:kilo-bhns}

In BHNS mergers, significant mass ejection and disk formation, leading to an electromagnetic counterpart, occur only if the black hole has a low mass and is rapidly spinning. In such cases, the neutron star undergoes tidal disruption during the final stages of the inspiral rather than being swallowed whole, which would otherwise result in negligible mass ejection.

A necessary condition for tidal disruption is that the neutron star’s tidal radius exceeds the innermost stable circular orbit (ISCO) of the black hole \citep{Metzger:2019zeh}. To determine whether a given BHNS system produces an electromagnetic counterpart, we adopt the model of \citet{Foucart:2018rjc}, where the remnant baryon mass is given by:
\begin{align}
\frac{m_{\rm rem}}{m_{\rm NS}^b} &= \left[ \max \left(\alpha \frac{1-2C_{\rm NS}}{\eta^{1/3}} - \beta \hat{R}_{\rm ISCO}\frac{C_{\rm NS}}{\eta}+\gamma,0\right)\right]^\delta \,, \label{eq:Mrem}  \\
C_{\rm NS} &= \frac{G m_{\rm NS}}{R_{\rm NS}c^2} \,,\\
\hat{R}_{\rm ISCO} &= 3+Z_2-{\rm sgn}(\chi_{\rm BH})\sqrt{(3-Z_1)(3+Z_1+2Z_2)}\,, \\
Z_1 &=1+\left(1-\chi_{\rm BH}^2 \right)^{1/3}\left[(1+\chi_{\rm BH})^{1/3}+(1-\chi_{\rm BH})^{1/3}\right] \,, \nonumber\\
Z_2 &=\sqrt{3\chi_{\rm BH}^2+Z_1^2} \,, \nonumber
\end{align}
where the ISCO radius is normalized as $\hat{R}_{\rm ISCO}=R_{\rm ISCO}/m_{\rm BH}$, and the model parameters are $\{ \alpha,\beta,\gamma,\delta\} = \{0.406, 0.139, 0.255,1.761 \}$. 

The neutron star compactness \( C_{\rm NS} \) depends on the neutron star radius, which is determined by the EOS. Here, we adopt a simplified assumption of $R_{\rm NS}=12.2$~km \citep{Rutherford:2024srk} and defer a more detailed analysis, considering EOS variations and their impact on the mass-radius relation, to future work. 

The baryonic mass $m_{\rm NS}^b$ exceeds the gravitational mass $m_{\rm NS}$ due to binding energy effects. A commonly used relation between the two, valid for non-rotating neutron stars, is
$m_{\rm NS}^b = m_{\rm NS} + A \, m_{\rm NS}^2$,
where $A=0.08$ \citep{Gao:2019vby}.

Given the adopted approximations and the uncertainties in the neutron star EOS, we use Eq.~\eqref{eq:Mrem} solely to verify whether the remnant baryon mass is nonzero. If $m_{\rm rem} > 0$, we assume the system produces an electromagnetic counterpart with a peak luminosity in the range \( 10^{41} - 10^{42} \) erg/s \citep{Metzger:2019zeh}, sampled uniformly in log-space.

\subsubsection{BNS}
\label{sec:kilo-bns}

\begin{figure}
\centering
\includegraphics[trim={0 0 0 0}, clip, width=1\columnwidth]{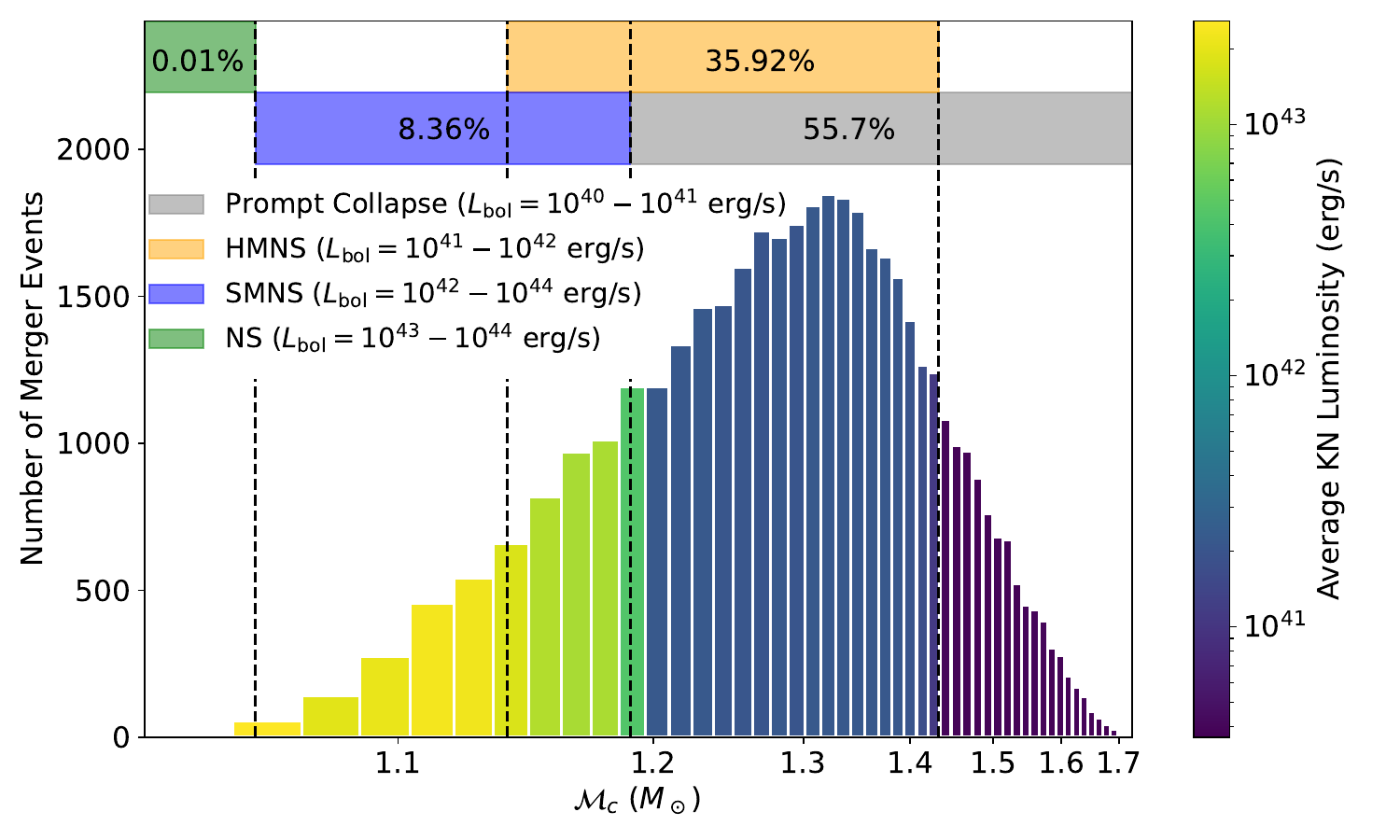}
\caption{The bands at the top indicate the chirp mass ranges associated with the four possible post-merger remnants, as described in the legend, along with their respective frequencies. The expected bolometric peak luminosity ranges for each remnant type are also shown.
The chirp mass distribution is derived from the component mass distributions in Section~\ref{NS-mass-dist}, while the color shading represents the average luminosity assigned to each chirp mass bin, following the procedure outlined in Section~\ref{sec:kilo-bns}.
}
\label{fig:KN_Luminosity}
\end{figure}

\citet{2019ApJ...880L..15M} introduce the Multi-Messenger (MM) matrix, a framework that links gravitational-wave and electromagnetic observables by correlating the chirp mass of a BNS system with its expected electromagnetic emission. The post-merger remnant is primarily determined by the component masses and can result in a stable neutron star (NS), a long-lived supermassive neutron star (SMNS), a short-lived hypermassive neutron star (HMNS), or prompt black hole formation (PC).

Due to uncertainties in the neutron star EOS, the mapping between chirp mass and remnant is not deterministic. Instead, multiple outcomes may arise for the same chirp mass \citep[][Table 1]{2019ApJ...880L..15M}. We sample merger outcomes according to this distribution, assuming equal probability within degenerate mass regions. The overlapping chirp mass ranges associated with different remnants are represented by the bands and their corresponding frequencies in Figure~\ref{fig:KN_Luminosity}.

Once merger remnants are assigned, kilonova luminosities are determined by sampling uniformly in log-space within the expected luminosity ranges also displayed in Figure~\ref{fig:KN_Luminosity} according to \citet[][Table 8]{Metzger:2019zeh}. This approach enables us to construct the kilonova luminosity distribution directly from the chirp mass distribution, which is inferred from the mass distributions in Section~\ref{NS-mass-dist}. The color shading in Figure~\ref{fig:KN_Luminosity} reflects the assigned luminosities.

The color scale in Figure~\ref{fig:KN_Luminosity} indicates the average peak bolometric luminosity within each chirp mass bin, revealing a clear trend where systems with lower chirp masses typically produce brighter kilonovae. This aligns with theoretical predictions, since mergers involving less massive neutron stars tend to eject larger amounts of material. The enhanced ejecta, combined with the amplification of strong magnetic fields, increases the electromagnetic output and leads to brighter kilonovae \citep{Metzger:2019zeh}. This feature has important implications for bright siren follow-up, particularly since gravitational-wave observations indicate that neutron stars in detected merging binaries are typically more massive than those in the Galactic population \citep{KAGRA:2021duu,2020RNAAS...4...65F}. Consequently, the BNS systems most easily detected via gravitational waves are also those expected to produce systematically fainter kilonovae.

This trend is expected, as lower-mass remnants are more likely to retain a significant fraction of the binary’s angular momentum, leading to stable and strong magnetic fields that enhance kilonova luminosity. In some cases, this process results in the formation of a highly magnetized neutron star, potentially a magnetar, which further amplifies the electromagnetic emission. In contrast, more massive remnants, such as hypermassive neutron stars (HMNS), typically collapse into black holes on short timescales (milliseconds), limiting the energy available for a bright kilonova.


\subsubsection{Band-Limited Luminosity}

During neutron star mergers, a substantial amount of ejecta (\(10^{-2} - 10^{-1} M_{\odot}\)) is expelled with velocities in the range \( v = 0.1 - 0.3 c \). Initially, this expanding material is highly opaque, preventing radiation from escaping. However, as the ejecta expands and becomes optically thin, radiation is released, leading to a peak in the luminosity curve. 
Numerical simulations indicate that this peak occurs at approximately \( t_{\text{peak}} \sim 3 \) days for mergers where the remnant undergoes a prompt collapse to a black hole. In contrast, for cases where the remnant forms a neutron star, the peak occurs at \( t_{\text{peak}} \lesssim 1 \) day. BHNS mergers, where the tidal ejecta component dominates, the peak is expected to occur at \( t_{\text{peak}} \sim 1 \) week \citep[][Table 8]{Metzger:2019zeh}.

Assuming that the emitted radiation is isotropic and follows a blackbody spectrum \citep{McCully:2017ljr}, the Stefan-Boltzmann law provides an estimate of the thermal emission temperature:
\begin{equation} \label{SB}
T = \left( \frac{L_{\text{bol}}}{4 \pi \sigma_{\text{SB}} R_{\text{mean}}^2} \right)^{1/4},
\end{equation}
where \(L_{\text{bol}}\) is the bolometric peak luminosity of the kilonova obtained in the previous sections, \(\sigma_{\text{SB}}\) is the Stefan-Boltzmann constant, and \(R_{\text{mean}}\) is the characteristic radius of the expanding ejecta at peak luminosity. 
To account for uncertainties in the ejecta expansion, we assume that \( R_{\text{mean}} \) follows a uniform distribution in log-space within the range \( [5\times 10^{14}, 5\times 10^{15}] \)~cm. This range is motivated by typical ejecta velocities and timescales associated with kilonova evolution.

The luminosity \( L_b \) in a specific observational band is obtained by weighting the blackbody spectral radiance \( B ( \nu,T) \) by the filter transmission function \( T_b(\nu) \), which defines the band’s observational response:
\begin{align}
L_b  &  =  L_{\text{bol}}  \frac{\pi \int_{0}^{\infty} B ( \nu,T) T_b(\nu)\, \d \nu}{\sigma_{\mathrm{SB}} T^4}\,, \label{L_band}\\
B ( \nu,T) &= \frac{2 h \nu^3}{c^2}  \frac{1}{\exp\left(\frac{h \nu}{k T}\right) - 1} \,.
\end{align}
The denominator in Eq.~\eqref{L_band} represents the total energy radiated per unit area by a blackbody over all frequencies, as given by the Stefan-Boltzmann law.
For \( T_b(\nu) \) we adopt the total transmission curves relative to the ugrizy LSST filters (\texttt{total\_*.dat}).%
\footnote{\href{https://github.com/lsst/throughputs/tree/main/baseline}{github.com/lsst/throughputs/tree/main/baseline}}
The absolute magnitude in a given band b is then obtained by comparing the object’s luminosity to the solar luminosity in the same band:
\begin{equation}
M_{b} = M_{\odot, b} - 2.5 \log_{10} \left( \frac{L_{b}}{L_{\odot, b}} \right) \,,
\end{equation}
where \( M_{\odot, b} \) is the absolute magnitude of the Sun in the same band, and \( L_{\odot, b} \) is the corresponding solar luminosity.

Following \citet{Hogg:1999ad}, the apparent magnitude in an observational band $b$ must include the redshift-dependent spectral correction:
\begin{align}
m_{b} &= M_{b} + 5\log_{10}\left(\frac{D_L}{10\,\mathrm{pc}}\right) + K_b(z) \,, \\
K_b(z) &= -2.5\log_{10}\left( (1+z)\frac{L_{b_z}}{L_{b}} \right) \,.
\label{m_nu}
\end{align}
The second term is the standard distance modulus, while the third term, \( K_b(z) \), represents the $k$-correction, accounting for spectral shifts caused by cosmological redshift. Specifically, the $k$-correction quantifies the difference between the observed redshifted luminosity $L_{b_z}$ and the luminosity $L_b$ in the object’s rest frame.
Explicitly, it is computed as:
\begin{align}
K_b(z) = -2.5\log_{10}\left( (1+z) \cdot \frac{ \int_0^\infty B\left((1+z)\nu, T\right) T_b(\nu)\, d\nu }{ \int_0^\infty B\left(\nu, T\right) T_b(\nu)\, d\nu } \right) \,.
\end{align}

\subsubsection{Kilonova VAC}

The value-added catalog includes bolometric and band-limited luminosities and magnitudes for all electromagnetic counterparts to BNS and BHNS mergers. This dataset enables multimessenger analyses by consistently connecting gravitational-wave detections and their associated kilonova emission to the corresponding host galaxies. The full structure of the \emph{Kilonova VAC} is presented in Table~\ref{tab:data-prod}.


\begin{table*}
\caption{Structure of the \texttt{CosmoDC2\_BCO} database, composed of input, output, and value-added catalogs, with representative examples.
\textbf{The top section} shows host galaxy parameters (redshift, sky location, stellar and halo masses, and LSST band magnitudes) drawn from the underlying \texttt{CosmoDC2} galaxy catalog.
\textbf{The second section} presents the \textit{Input Catalog}, listing merger type (BBH, BNS, BHNS), intrinsic and extrinsic parameters, and assigned simulation tile. The complete catalog also contains several derived quantities. Spins for BNS mergers are fixed to zero due to waveform model constraints, and coalescence times are conventionally set to zero.
\textbf{The third section} provides the CSV-formatted \textit{Output Catalog}, reporting individual detector and network signal-to-noise ratios ($\rho$), marginalized parameter uncertainties derived from Fisher matrices, and flags indicating successful parameter estimation (invertible Fisher matrices). Undetected events or those yielding non-invertible matrices are marked with NaN entries. The FITS version of this catalog additionally contains the complete Fisher and covariance matrices.
\textbf{The bottom section} describes the \textit{Kilonova Value-Added Catalog}, detailing kilonova classification, bolometric luminosities, and photometric data. Remnant types for BNS mergers are classified following criteria discussed in Section~\ref{sec:kilo-bns}; BHNS mergers are categorized (bright or dark) based on neutron star tidal disruption, as outlined in Section~\ref{sec:kilo-bhns}. Examples illustrate diverse outcomes, including a promptly collapsing BNS merger and a BHNS event without tidal disruption.
All angular parameters are given in radians, except RA and DEC; masses are in solar masses, and luminosity distances in Gpc.
\label{tab:data-prod}}
\centering
\setlength{\tabcolsep}{9pt}
\renewcommand{\arraystretch}{1.2}
\begin{tabular}{cccccc}
\multicolumn{6}{c}{\bf CosmoDC2} \\
\toprule
Galaxy\_ID & $z$ & RA/DEC &  Stellar mass & Host-halo mass & $ugrizy$-band vector \\
\midrule
3004438004 & 1.37 & 73.0º, -28.2º   & $5.4 \times 10^{10}$ & $1.7 \times 10^{13}$ & $\{ 25.2, \ldots, 22.5  \}$ \\
\ldots & \ldots & \ldots & \ldots & \ldots\\
\bottomrule
\noalign{\vskip 3ex}
\multicolumn{6}{c}{\bf Input Catalog} \\
\toprule
Galaxy\_ID & Tile & BCO & $z$ & Intrinsic parameters $\theta_{\rm gw}^{\rm int}$ of Eq.~\eqref{15int}  & Extrinsic parameters $\theta_{\rm gw}^{\rm ext}$  of Eq.~\eqref{15ext}\\
\midrule
3004438004 & 0  & BBH & 1.37& $\{27, 0.25, 0.11, 2.9, 3.8, 0.23, 1.3, 2.3 \}$ & $\{ 9.6, 73.0, -28.2, 1.37, 0.80, 0, 3.3    \}$ \\
3452407336 & 0  & BNS & 2.29& $\{4.95, 0.25, 0.035, 0, 0, 0.060, 0, 0 \}$ & $\{ 17.9, 59.8, -30.5, 1.94, 1.2, 0, 0.68   \}$ \\
1827406177 & 0  & BHNS& 2.30 & $\{20.5, 0.034, 0.2, 0.6, 2.1, 0.005, 1.1, 2 \}$ & $\{ 18, 50.1, -27.3, 2.4, 2.4, 0, 1.3  \}$ \\
\ldots & \ldots & \ldots & \ldots & \ldots\\
\bottomrule
\noalign{\vskip 3ex}
\multicolumn{6}{c}{\bf Output Catalog} \\
\toprule
Galaxy\_ID & Tile & BCO & $\{ \rho_i \}, \rho_{\rm net}$ & Marginalized errors on $\theta_{\rm gw}$ of Eq.~\eqref{15par}  & Fisher matrix flag \\
\midrule
3004438004 & 0& BBH & $\{  77,  \ldots,  0.6 \}$, 78 & $\{ 0.00073, \ldots,  2.6 \}$ & True \\
3452407336 & 0 & BNS& $\{  8.4,  \ldots,  0.07 \}$, 8.6 & NaN & NaN \\
1827406177 & 0& BHNS & $\{  27,  \ldots,  0.18 \}$, 28 & $\{ 0.0057, \ldots,  1.7 \}$ & True \\
\ldots & \ldots & \ldots & \ldots & \ldots\\
\bottomrule
\noalign{\vskip 3ex}
\multicolumn{6}{c}{\bf Kilonova Value-Added Catalog} \\
\toprule
Galaxy\_ID & Tile & BCO & Remnant type & Luminosity & $ugrizy$-band  vector \\
\midrule
3452407336 & 0 & BNS& PC & $9 \times 10^{40} \frac{\rm erg}{\rm s}$ & $\{ 40.1, \ldots, 37.1  \}$ \\
1827406177 & 0 & BHNS& dark & 0 & NaN \\
\ldots & \ldots & \ldots & \ldots & \ldots\\
\bottomrule
\end{tabular}
\end{table*}

\section{Results}
\label{results}

\begin{figure}
\centering 
\includegraphics[trim={0 0 0 0}, clip, width= 1 \columnwidth]{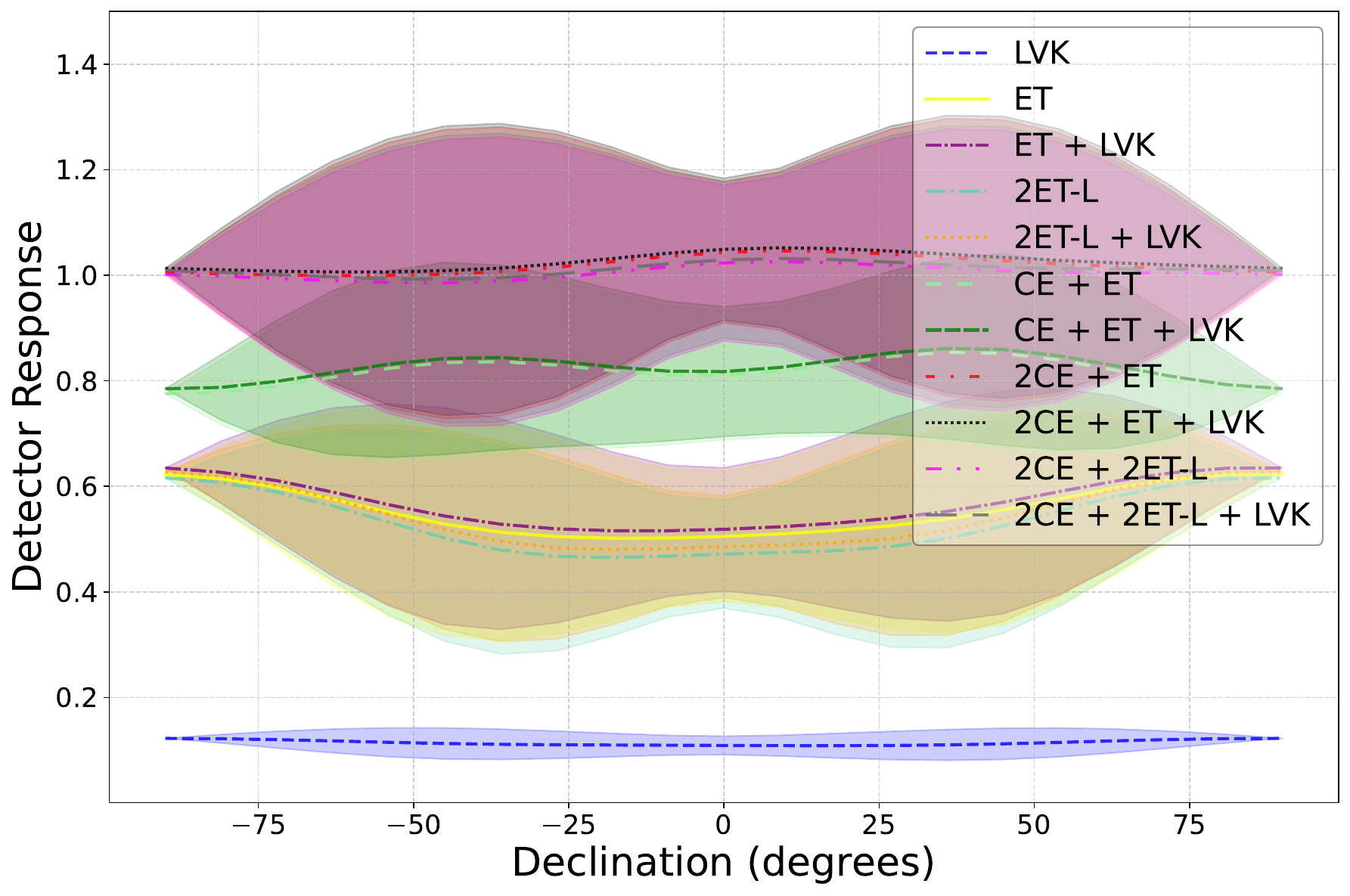}
\caption{Declination-dependent sensitivity of the detector network, computed as the effective quadratic sum of pattern functions, weighted by the inverse of the noise power spectrum. The solid line represents the mean response across all right ascensions, while the shaded region indicates the standard deviation due to variation in right ascension. See Section~\ref{res:net} for details.}
\label{response}
\end{figure}

We present the gravitational-wave and electromagnetic results obtained from the synthetic catalogs generated via the pipeline described in Section~\ref{section_pipeline}. These results characterize the expected performance of future gravitational-wave detector networks and their synergy with electromagnetic follow-up programs.

A key feature of this work is the consistent association of synthetic GW and EM events with realistic host galaxies from the \texttt{CosmoDC2} catalog. Table~\ref{tab:data-prod} summarizes the structure of our \texttt{CosmoDC2\_BCO} database, which comprise three sets of catalogs designed for integration with the \texttt{CosmoDC2} galaxy catalog. All catalogs consistently share the same \texttt{Galaxy\_ID}, enabling seamless crossmatching with the more than 500 properties describing each host galaxy, including observational systematics.

The input catalog assigns, via the pipeline shown in Figure~\ref{fig:flowchart}, a set of intrinsic and extrinsic gravitational-wave parameters to each host galaxy. The output catalog provides the network SNR and, if the event is detected, the associated parameter covariance matrix. The output catalogs are available for different detector networks, waveform models, and duty cycle assumptions. In the CSV format, only the marginalized uncertainties (diagonal of the covariance matrix) are included, while the FITS version contains the full Fisher and covariance matrices.
The full set of data products is publicly available at \href{https://github.com/LSSTDESC/CosmoDC2_BCO}{github.com/LSSTDESC/CosmoDC2\_BCO}. This dataset is well suited for detailed gravitational-wave analyses and cosmological inference using dark and bright sirens, including effects from large-scale structure, which will be investigated in future work.

We consider a comprehensive set of network configurations, as shown in Table~\ref{tab:detections_networks}, to assess the impact of future 3G detector designs. In particular, we compare the triangular ET with the alternative 2ET-L configuration and evaluate the continued relevance of the current 2G network. A key question is how long existing 2G detectors should remain operational alongside 3G observatories. The following sections present forecasted constraints, systematically comparing these different network configurations.

For our analysis, we primarily use results from \texttt{tile~\#0}, covering \sky~deg$^2$, except for kilonova detection, which is analyzed across all 36 tiles due to the limited number of expected events. However, the full dataset is available in our GitHub repository.
The structure of the results is as follows: 
\begin{itemize}
\item Section~\ref{res:net} discusses the network response.
\item Section~\ref{res:det} presents the expected number of detections.
\item Sections~\ref{res:astro} and~\ref{res:cosmo} analyze the uncertainties on  astrophysical and cosmological parameters, respectively.

\item Section~\ref{res:correlations} analyzes the correlations between the GW parameters, the SNR, and the redshift.
\item Section~\ref{res:dc} examines the effect of duty cycle limitations.
\item Section~\ref{res:kn} examines the abundance of kilonovae.
\end{itemize}
Additional analyses are provided in the appendices:
\begin{itemize}
\item Validation of the Fisher approximation via full MCMC is discussed in Appendix~\ref{ap:validation}.
\item Waveform model comparisons are presented in Appendix~\ref{ap:approximant}.
\item Further analyses available in the repository are detailed in Appendix~\ref{ap:further}.
\end{itemize}

As shown in Table~\ref{tab:data-prod}, a flag indicates whether the Fisher matrix is numerically invertible, as defined in Section~\ref{fisher}. The results discussed in the following sections are restricted to events with successful inversion. While this selection may introduce bias, an analysis of all catalogs produced to date—including various setups and sky tiles—shows that the impact of the inversion requirement is generally limited. On average, fewer than 1\% of detected events are excluded, although in extreme cases, particularly for the lowest duty cycles, the fraction can rise to 10\%.

\subsection{Network Response}
\label{res:net}

The sensitivity of a gravitational-wave detector network varies across the sky due to the response of individual detectors and their relative orientations. To quantify this, we compute the effective network response as the quadratic sum of the pattern functions of individual detectors, weighted by the inverse of their respective noise power spectra, as defined in Eq.~\eqref{Fnet}. 

Figure~\ref{response} illustrates the network response as a function of declination, averaged over all right ascensions. The shaded regions represent the standard deviation due to right ascension variations, highlighting the anisotropic sensitivity of the network. The response is normalized relative to the average response of the CE detector.
The addition of LVK to ET improves the uniformity of the response across the sky, but its impact becomes negligible once CE is included.

For a complete visualization of the network response across the full sky, including right ascension dependence, we provide full-resolution RA-DEC maps for all network configurations in our \href{https://github.com/LSSTDESC/CosmoDC2_BCO}{GitHub repository}.

\subsection{Number of detections}
\label{res:det}

\begin{figure*}
\centering 
\includegraphics[trim={0 0 0 0}, clip, width= 1 \textwidth]{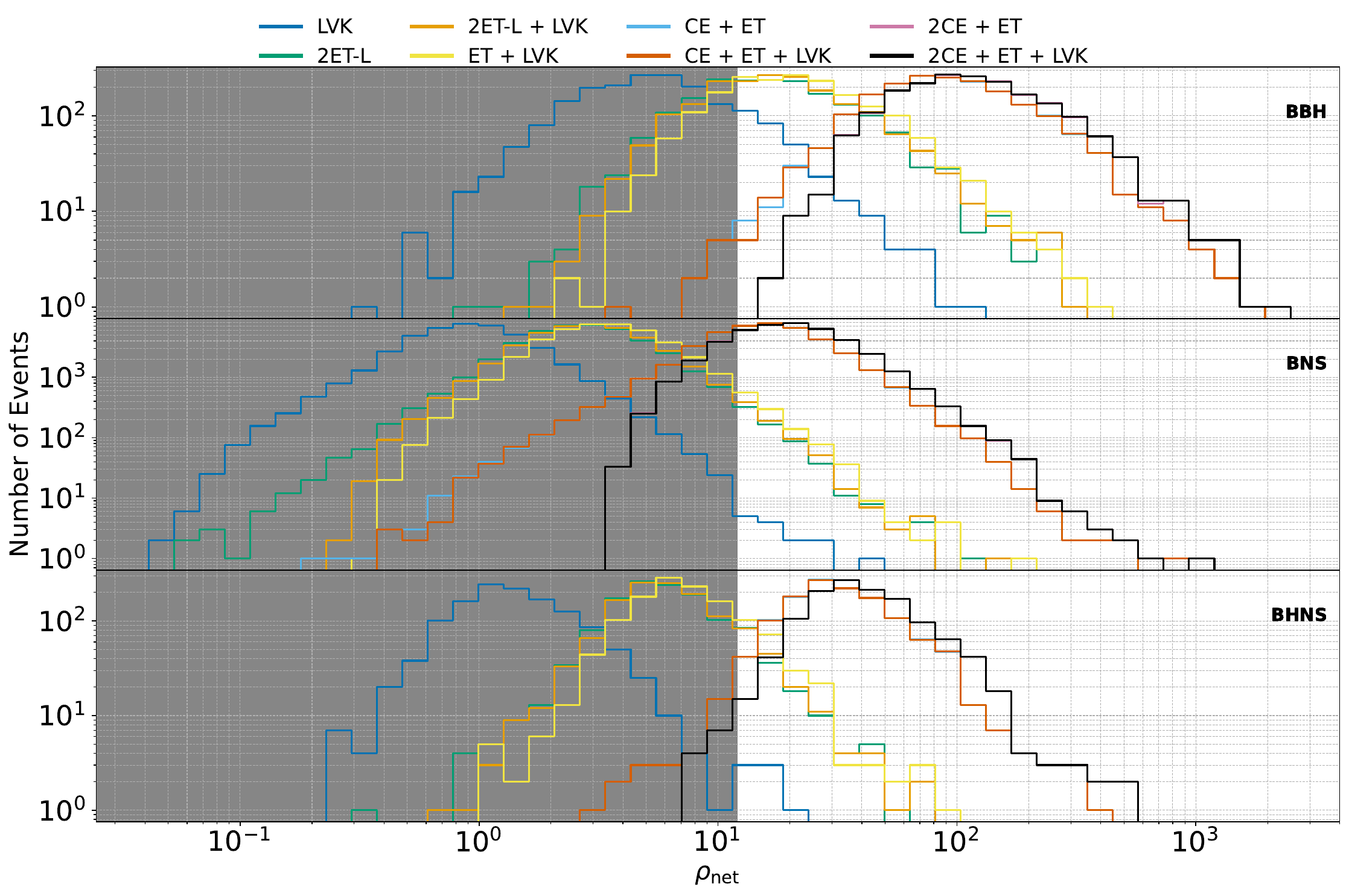}
\caption{SNR distribution for BBH (top), BNS (middle), and BHNS (bottom) mergers from \texttt{tile~\#0}, assuming a 100\% duty cycle.
The gray shaded area indicates the region below the detection threshold of 12. The total number of  events is identical across all network configurations.}
\label{SNRs_net}
\end{figure*}

Here we discuss the number of detections for each gravitational-wave network configuration. These numbers are slightly underestimated due to the redshift limitation of the \texttt{CosmoDC2} catalog, which includes galaxies only up to $z = 3$, whereas our merger rate model extends to $z = 7$. The last two rows of Table~\ref{tab:detections_networks} report the total number of mergers for $z < 7$ and $z < 3$, respectively. All results presented in the following sections are restricted to events within $z < 3$.

Unless otherwise stated, we consider only the catalogs with a $100\%$ duty cycle to isolate and assess the intrinsic capabilities of the gravitational-wave networks and the physical properties of the BCO population. In Section~\ref{res:dc}, we present results for more realistic scenarios that account for non-ideal duty cycle values.

Figure~\ref{SNRs_net} presents the distribution of BBH (top), BNS (middle), and BHNS (bottom) mergers from \texttt{tile~\#0} as a function of the network SNR $\rho_{\rm net}$ defined in Eq.~\eqref{rhonet}, for the eight detector network configurations of Table~\ref{tab:detections_networks}. Results employ waveform models \texttt{IMRPhenomXPHM} for BBH and BHNS systems and \texttt{IMRPhenomXHM} for BNS.
Only events with \( \rho_{\rm net} \geq 12 \) are considered detectable; the gray region marks the undetectable SNR range. We find an overall agreement with the previous results by \citet{Colombo:2023une,Loffredo:2024gmx}.

The detection fraction varies significantly across network configurations; Table~\ref{tab:detections_networks} reports the corresponding event counts assuming a 70\% duty cycle.
Detection efficiency also depends on the physical complexity of the adopted waveform model, as richer physical models typically yield higher SNR values; this relationship is examined in detail in Appendix~\ref{ap:approximant}.

\begin{table}
\centering
\caption{Number of detected events and corresponding percentages (in parentheses) within $z<3$ for each binary type from \texttt{tile~\#0} (\sky deg$^2$), assuming a 70\% duty cycle and 10 years of observations. The results refer to the gravitational-wave network configurations considered in this study. The adopted waveform approximants were \texttt{IMRPhenomXPHM} for BBH and BHNS systems, and \texttt{IMRPhenomXHM} for BNS systems.}
\label{tab:detections_networks}
\setlength{\tabcolsep}{5pt}
\renewcommand{\arraystretch}{1.2}
\begin{tabular}{lccc}
\toprule
Network & BBH & BNS & BHNS \\
\midrule
LVK & 184 (9.7\%) & 10 (0.02\%) & 2 (0.16\%) \\
2ET-L & 935 (49.4\%) &358 ( 0.82\%) & 87 (6.9\%) \\
\midrule
2ET-L+LVK & 1008 (53.3\%) & 402 ( 0.92\%) & 99 (7.8\%) \\
ET+LVK & 1053 ( 55.66\%) & 654 ( 1.5\%) & 144 (11.4\%) \\
\midrule
CE+ET & 1621 (85.7\%) & 20558 (47.2\%) & 867 (68.3\%) \\
CE+ET+LVK & 1627 (86.0\%) & 20683 (47.5\%) & 864 (68.1\%)\\
\midrule
2CE+ET & 1809 (95.6\%) & 28007 (64.3\%) &  1144 (90.2\%) \\
2CE+ET+LVK & 1810 (95.7\%) & 27966 (64.2\%) & 1147 (90.4\%) \\
\midrule
All Mergers ($z<3$) & 1892  & 43577  & 1269  \\
All Mergers ($z<7$) & 2200  & 49530  & 1830  \\
\bottomrule
\end{tabular}
\end{table}

We observe that the distributions in Figure~\ref{SNRs_net} cluster according to the number of 3G detectors included in the network configuration. The addition of the Einstein Telescope leads to an approximate fourfold increase in the signal-to-noise ratio, with an additional factor of approximately 4–5 achieved by including Cosmic Explorer. Adding a second CE detector provides a more modest improvement of about 25\%. This diminishing return suggests that further substantial enhancements in network sensitivity may only be attainable with future, beyond-3G detector concepts.

Comparing the triangular ET configuration with the alternative 2ET-L arrangement, assuming the LVK detectors remain operational, we find comparable performance for BBH detections. However, the triangular ET setup, which includes an additional detector, results in approximately 50\% more BNS and BHNS detections, as shown in Table~\ref{tab:detections_networks}.
Finally, the continued operation of 2G detectors provides a clear benefit only when used in conjunction with the triangular ET alone.

\subsubsection*{Comparison with O4}

As of March 20, 2025,%
\footnote{\url{https://www.ligo.caltech.edu/LA/news/ligo20250320}} the LVK network has reported 290 gravitational-wave detections from four observing runs spanning a total of 46 months. Among these, 2--3 are classified as BNS mergers and 5--6 as BHNS candidates.
Focusing on the fourth observing run (O4)—the most relevant for comparison with the O5-like sensitivities adopted in this work—200 events were detected during its 20 months of operation.
During the 8 months of O4a, the LIGO detectors did not achieve the expected O4 design sensitivity, and Virgo was not operational. In the subsequent 12 months of O4b, LIGO improved its sensitivity but still fell short of the nominal O4 design goals. Virgo reached a sensitivity only marginally better than its O3 performance. KAGRA did not attain competitive sensitivity throughout O4. Out of the 200 gravitational-wave candidates reported during this run, only 2 were classified as BNS and 3 as BHNS merger candidates.
The duty cycles during O4, including both O4a and O4b, were 65\% for LIGO Hanford, 71\% for LIGO Livingston, and 53\% for coincident operation of both LIGO detectors \citep{Capote:2024rmo}.

Table~\ref{tab:detections_networks} summarizes our forecast for the upcoming O5 observing run with the fully operational LVK network. Assuming a 70\% duty cycle (higher than the actual duty cycle achieved in O4), full-sky coverage (approximately 94 times the area of \texttt{tile~\#0}), and 20 months of observations, we anticipate approximately 3000 BBH
and 150 BNS
detections—a substantial increase compared to O4. For comparison, using only the LIGO network at its O4 design sensitivity (again higher than actually achieved) and assuming a 60\% duty cycle, we would forecast approximately 550 BBH
and 8 BNS
detections over the same observation period and full-sky coverage. Considering the inherent differences between idealized forecasts and real-world operations and the uncertainties on the merger rates, these estimates align closely with the actual order-of-magnitude number of detections observed in O4 and forecasted in \citet{Colombo:2025sdm}.

\subsection{Uncertainty on  astrophysical parameters}
\label{res:astro}

\begin{figure*}
\centering 
\includegraphics[trim={0 0 0 0}, clip, width= 1 \textwidth]{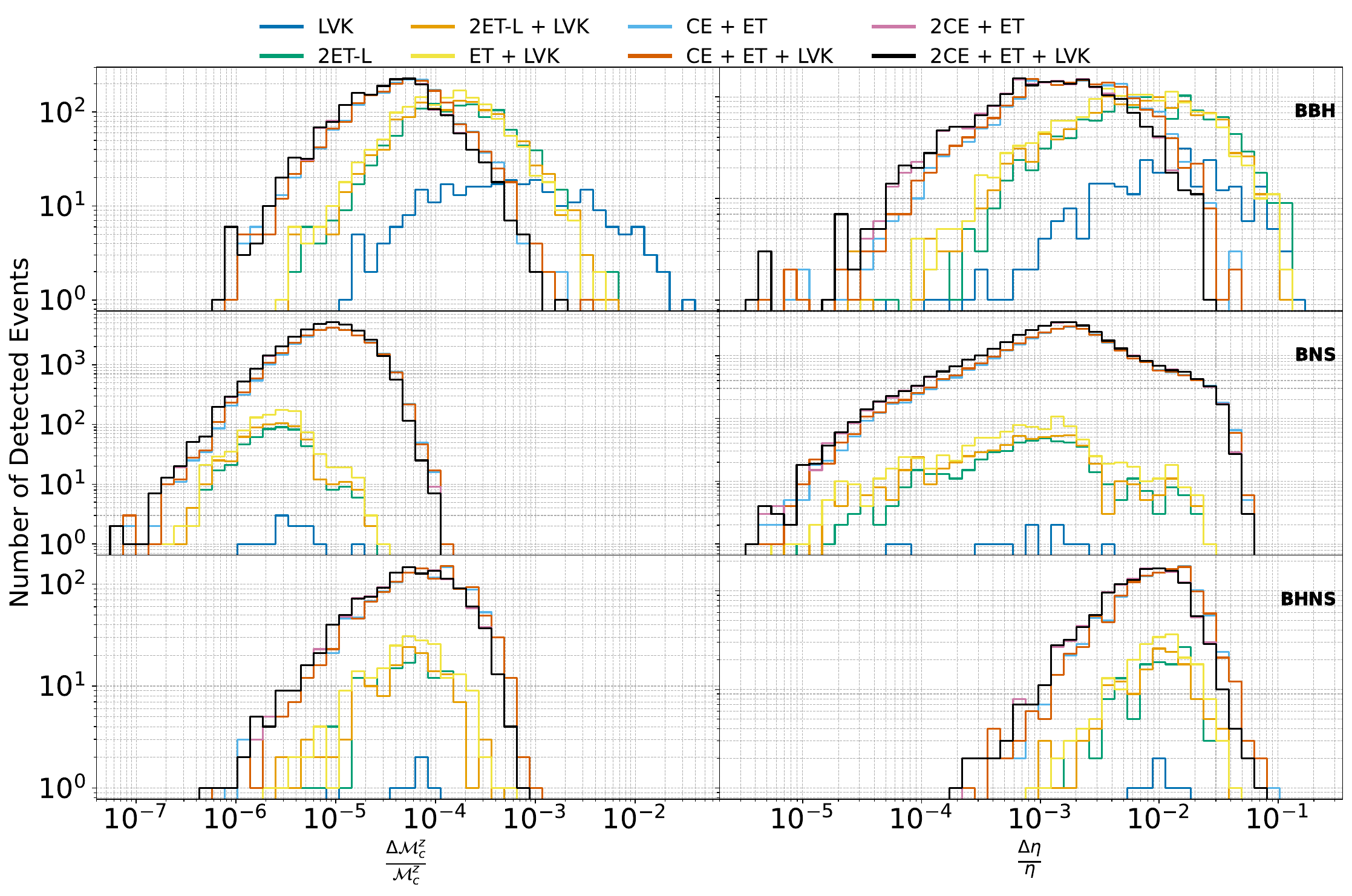}
\caption{Relative uncertainties on redshifted chirp mass (left) and symmetric mass ratio (right) for BBH (top), BNS (middle), and BHNS (bottom) mergers from \texttt{tile~\#0}. Results assume a 100\% duty cycle. }
\label{Mc-eta_err_net}
\end{figure*}

Figure~\ref{Mc-eta_err_net} shows the distribution of BBH (top), BNS (middle), and BHNS (bottom) mergers from \texttt{tile~\#0} as a function of the relative uncertainty on the redshifted chirp mass (left) and the symmetric mass ratio (right).
In this and the following sections, we assume a 100\% duty cycle and adopt the waveform models \texttt{IMRPhenomXPHM} for BBH and BHNS systems, and \texttt{IMRPhenomXHM} for BNS. Only events with a reliable covariance matrix are included, as discussed in Section~\ref{fisher}.

Since only detected events are included, superior network configurations naturally yield more populated histograms.
The Einstein Telescope notably reduces the higher-uncertainty tail seen for BBH mergers with the LVK network. While more advanced networks detect a higher fraction of events with precisely constrained parameters (left tails), their enhanced detection sensitivity also includes events with inherently larger uncertainties (right tails), as in the case of BNS and BHNS mergers. 

The distributions clearly separate into three groups: LVK alone, networks incorporating ET (with minimal differences between the two ET configurations), and networks that include CE.
The smaller uncertainties observed for BNS mergers are primarily due to their typically higher SNR--resulting from their longer inspiral phase--combined with their lower chirp mass and closer distances to the observer, all of which contribute to improved parameter estimation precision.
It is noteworthy that these parameter uncertainties depend strongly on the choice of waveform model; see Appendix~\ref{ap:approximant}, where we show that including higher-order modes yields a substantial improvement, whereas spin precession produces only a modest change.

\begin{figure*}
\centering 
\includegraphics[trim={0 0 0 0}, clip, width= 1 \textwidth]{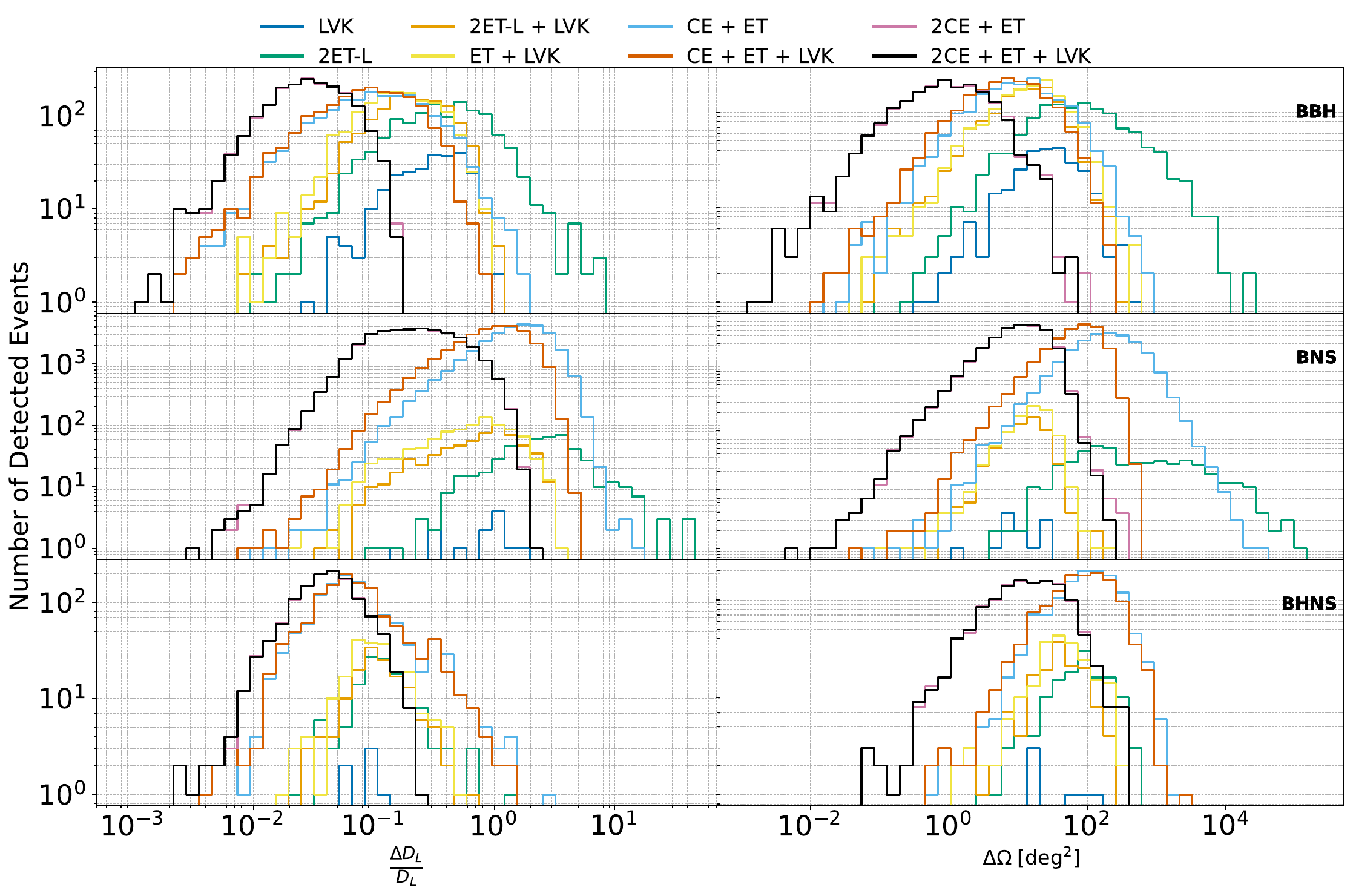}
\caption{Distributions of BBH (top), BNS (middle), and BHNS (bottom) mergers from \texttt{tile~\#0}, showing the relative uncertainty on luminosity distance (left) and the 95\% confidence-level sky localization area (right). Results assume a 100\% duty cycle.}
\label{DL-omega_err_net}
\end{figure*}

\subsection{Uncertainty on  cosmological parameters}
\label{res:cosmo}

Figure~\ref{DL-omega_err_net} presents the distributions of BBH (top), BNS (middle), and BHNS (bottom) mergers from \texttt{tile~\#0} as a function of the relative uncertainty on the luminosity distance $D_L$ (left) and the 95\% confidence-level sky localization area $\Delta \Omega$ (right). These parameters are critical for dark siren cosmology and electromagnetic follow-up observations of bright sirens.

The addition of more detectors to the gravitational-wave network significantly improves the constraints on both the luminosity distance $D_L$ and the sky localization area $\Delta \Omega$, notably resolving the bimodality in sky position that typically affects networks with only two detectors.
This enhancement is more pronounced than the improvements seen for the redshifted chirp mass and the symmetric mass ratio, discussed in Section~\ref{res:astro}.
Particularly striking is the beneficial impact of 2G detectors, which substantially enhance sky localization due to improved triangulation and more uniform sky coverage. Although 2G detectors might not significantly contribute to increasing the detection rate (see Figure~\ref{SNRs_net}), their presence is crucial for precise distance estimation and sky localization, thereby greatly reducing the number of potential host galaxies in dark siren cosmological analyses, especially for BBH systems.

For BNS mergers, the importance of retaining 2G detectors alongside 3G instruments is particularly evident. The 2ET-L configuration alone yields significantly poorer constraints on $D_L$ and $\Delta \Omega$ compared to networks incorporating 2G detectors. (We do not show results for the single-location triangular ET alone, given its inherently limited localization capability.) As a result, the continued operation of 2G detectors will remain essential for effective electromagnetic follow-up. Their contribution becomes negligible only once a network of three 3G detectors is operational, in agreement with \citet{Ferri:2024amc}. 
Comparing the triangular ET configuration with the alternative 2ET-L arrangement, when combined with the operational LVK network, we find that their performances are broadly comparable.

For BHNS binaries, the improvement provided by including 2G detectors is less pronounced. This behavior is further illustrated in Figure~\ref{fig:iota_uncertainties}, which shows the distribution of the relative uncertainty on the inclination angle $\iota$, a parameter strongly degenerate with luminosity distance $D_L$. Improvements in the constraints on $\iota$ are notably greater for BBH and BNS systems compared to BHNS, corroborating the larger enhancements observed in the uncertainty of $D_L$ for these two classes.
The distinct behavior of BHNS mergers may be attributed to their smaller mass ratios, which affect both the signal morphology and the parameter estimation accuracy; see Figure~\ref{fig:mass-dist}.

\begin{figure}
\centering 
\includegraphics[trim={0 0 0 0}, clip, width= 1 \columnwidth]{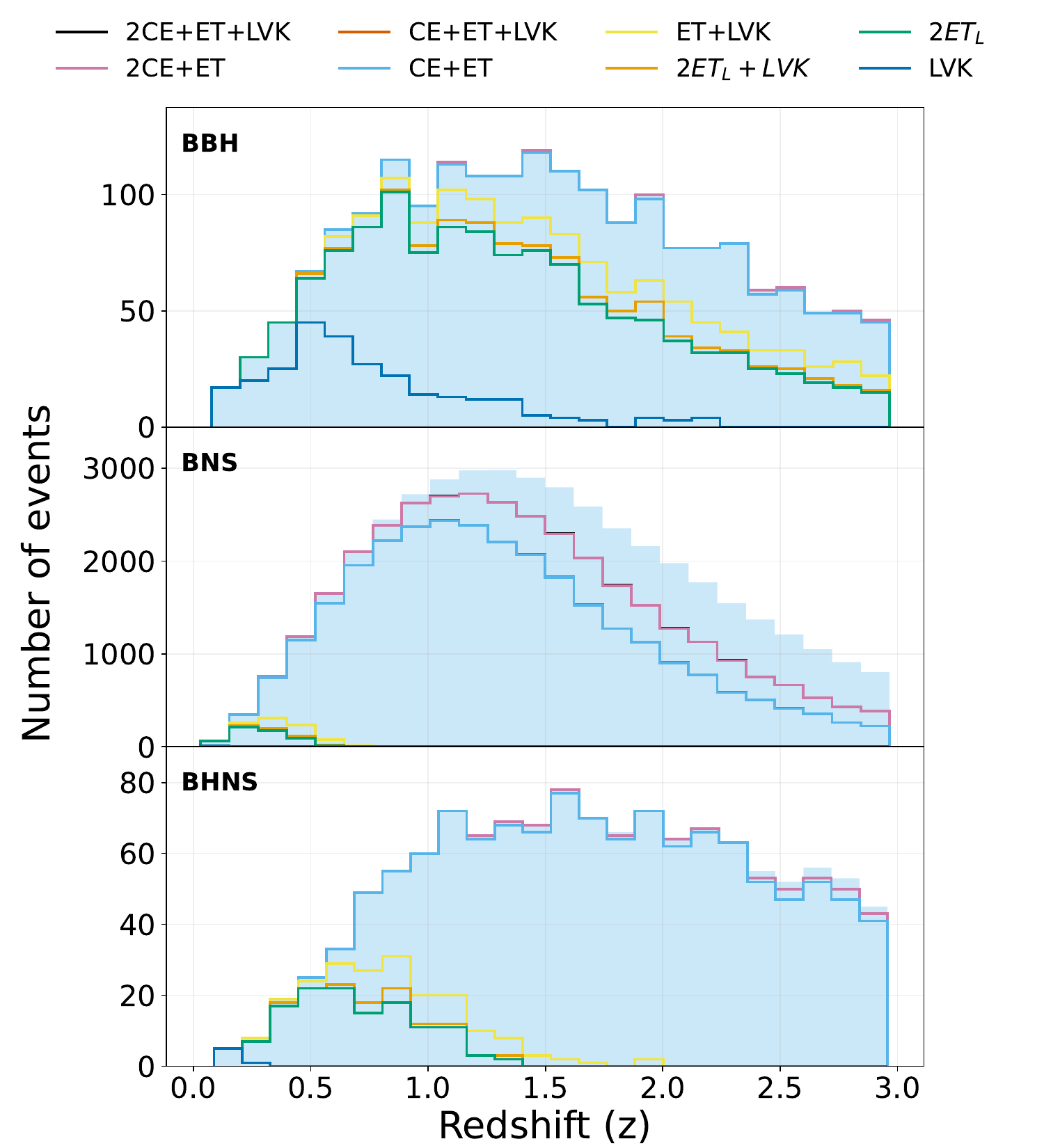}
\caption{Host-galaxy redshift distributions for BBH (top), BNS (middle), and BHNS (bottom) mergers in \texttt{tile~\#0}, assuming different detector networks operating with a 100\% duty cycle. The filled blue histogram represents the total number of merger events.
}
\label{fig:zplot}
\end{figure}

Finally, Figure~\ref{fig:zplot} shows the distribution of host-galaxy redshifts for the three types of coalescences, explicitly illustrating the number of events missed due to the redshift limitation imposed by the \texttt{CosmoDC2} simulations. This selection effect does not significantly affect the detection rates for the LVK and ET+LVK configurations in the case of BNS and BHNS sources. For BBH systems, the detection rates are slightly underestimated for the CE+ET+LVK and ET+LVK configurations, with a smaller impact for the latter, while the LVK network remains essentially unaffected. These different biases across network configurations are expected, since more sensitive networks probe larger distances and are therefore more impacted by the redshift limitation intrinsic to the background galaxy catalog. Quantitatively, this effect leads to an expected reduction in detection counts of at most $\sim10\%$ for BBH and BNS systems, and up to $\sim20\%$ for BHNS systems (see Table~\ref{tab:detections_networks}).

In practice, these estimates may be conservative, as detectability continues to decrease beyond the imposed redshift limit.
From the perspective of scientific usability, particularly for cosmological applications, the impact of these missing high-redshift events is even less significant, as the excluded events are expected to have larger relative uncertainties in their inferred parameters, which reduces their statistical weight in cosmological analyses.

\subsection{Correlations}
\label{res:correlations}

\begin{figure*}
\centering 
\includegraphics[trim={0 0 0 0}, clip, width= 1\textwidth ]{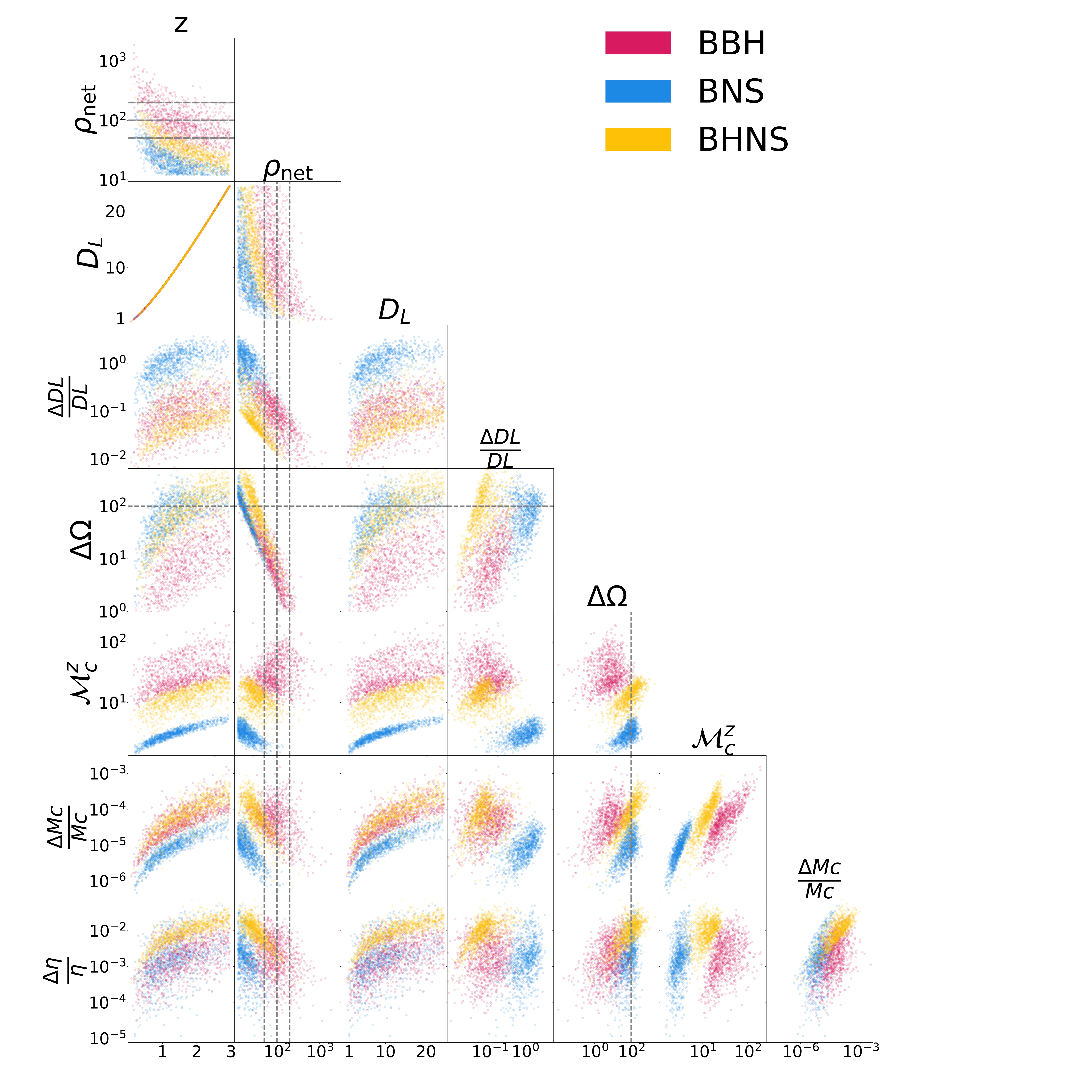}
\caption{Correlations among redshift, network SNR ($\rho_{\rm net}$), luminosity distance ($D_L$, in Gpc), sky localization area ($\Delta \Omega$, in deg$^2$), redshifted chirp mass ($\mathcal{M}_c^z$, in solar masses), and fractional uncertainties for luminosity distance, chirp mass, and symmetric mass ratio ($\eta$). The panels show scatter plots for BBH, BNS, and BHNS mergers detected by the CE+ET+LVK network with 100\%\ duty cycle. Dashed lines indicate SNR values of 50, 100, and 200, and $\Delta \Omega =100$~deg$^2$. Parameter estimation accuracy depends on source properties and SNR, with distinct trends observed for different merger types.}
\label{corr_triplot}
\end{figure*}

Figure~\ref{corr_triplot} illustrates the correlations among several key parameters, including redshift, network SNR, luminosity distance, sky localization area, redshifted chirp mass, and their fractional uncertainties. The distributions for BBH, BNS, and BHNS mergers show clear distinctions due to differences in intrinsic parameters and detection characteristics. Results correspond to the CE+ET+LVK network with 100\% duty cycle.

Dashed lines at $\rho_{\mathrm{net}}=50$, 100, and 200 help to highlight the subset of high SNR events, which are particularly relevant for precision cosmology. In the first column a clear trend is observed in the form of a power-law-like decrease of the network SNR with redshift.  This behavior is consistent across all BCO types in our simulated catalog. BBH systems populate the highest SNR regime, followed by BHNS and then BNS. This ordering arises because the larger chirp masses of BBH produce intrinsically stronger amplitudes, which dominate over the longer inspiral durations of lighter systems. However, this clear separation in SNR can be reduced or even inverted in incomplete samples from less sensitive networks.

The second column highlights the anti-correlation between SNR and parameter uncertainties. Among these, the sky localization area ($\Delta \Omega$) shows one of the strongest dependencies, with higher SNR events being localized much more precisely. Additionally, the third column reveals a positive correlation between the fractional uncertainty on luminosity distance ($\Delta D_L / D_L$) and the sky localization area. This relationship likely occurs because a poorly constrained sky position prevents the full disentanglement of the signal's polarization content, a degeneracy that propagates into the distance measurement.

Finally, the rightmost panels emphasize the correlations within the mass sector. Across all BCO classes, larger uncertainties in the chirp mass $\mathcal{M}_c$ are accompanied by larger uncertainties in the symmetric mass ratio $\eta$. This trend is expected since both parameters contribute to the phasing of the inspiral waveform and can partially compensate one another in the estimation process. As a result, statistical errors in one parameter tend to propagate into the other, broadening their joint posterior distributions.

\subsection{Impact of Duty Cycle}
\label{res:dc}

\begin{figure}
\centering 
\includegraphics[trim={0 0 0 0.7cm}, clip, width= 1 \columnwidth]{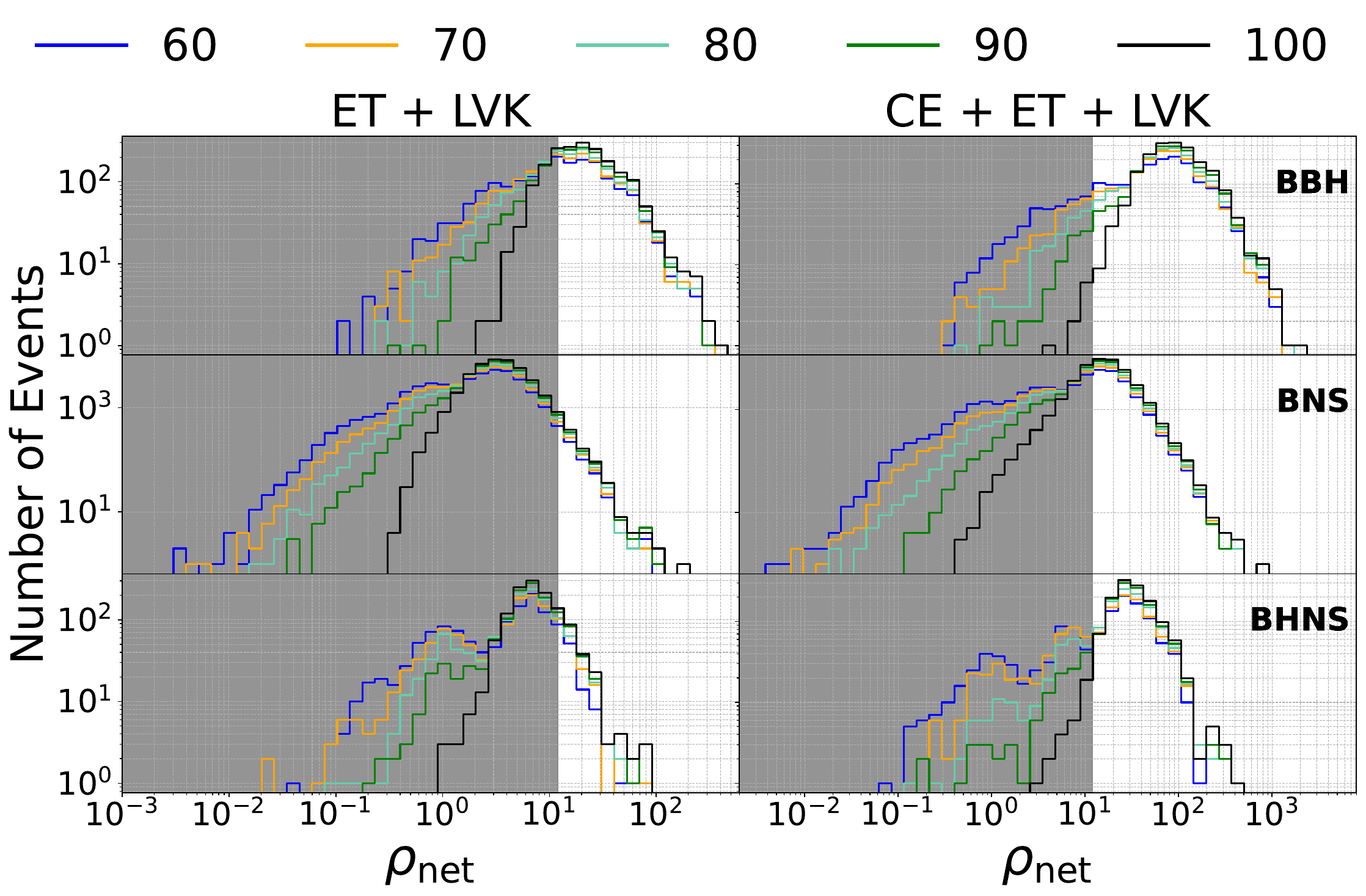}
\includegraphics[trim={0 0 0 0.7cm}, clip, width= 1 \columnwidth]{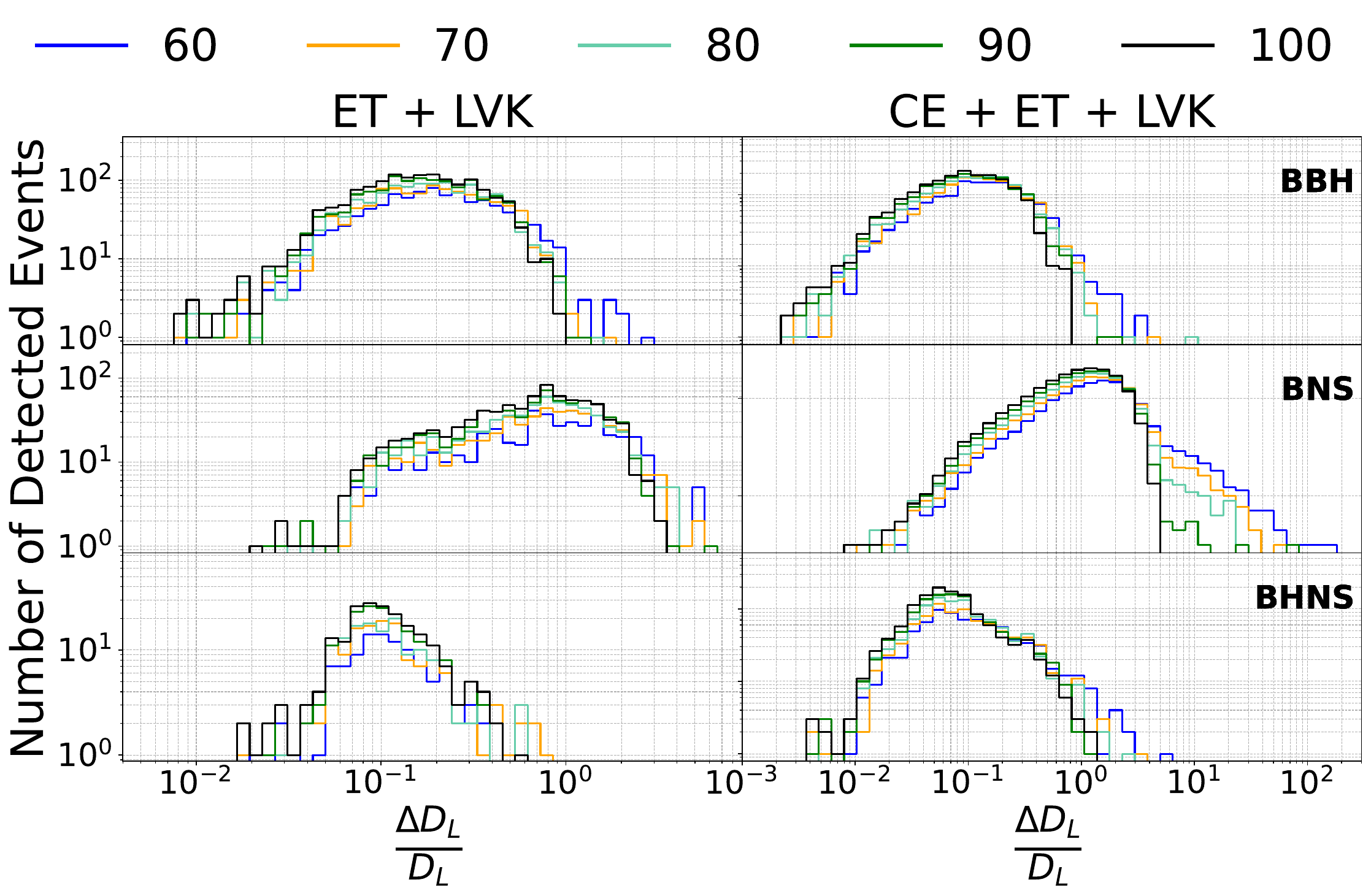}
\includegraphics[trim={0 0 0 0.7cm}, clip, width= 1 \columnwidth]{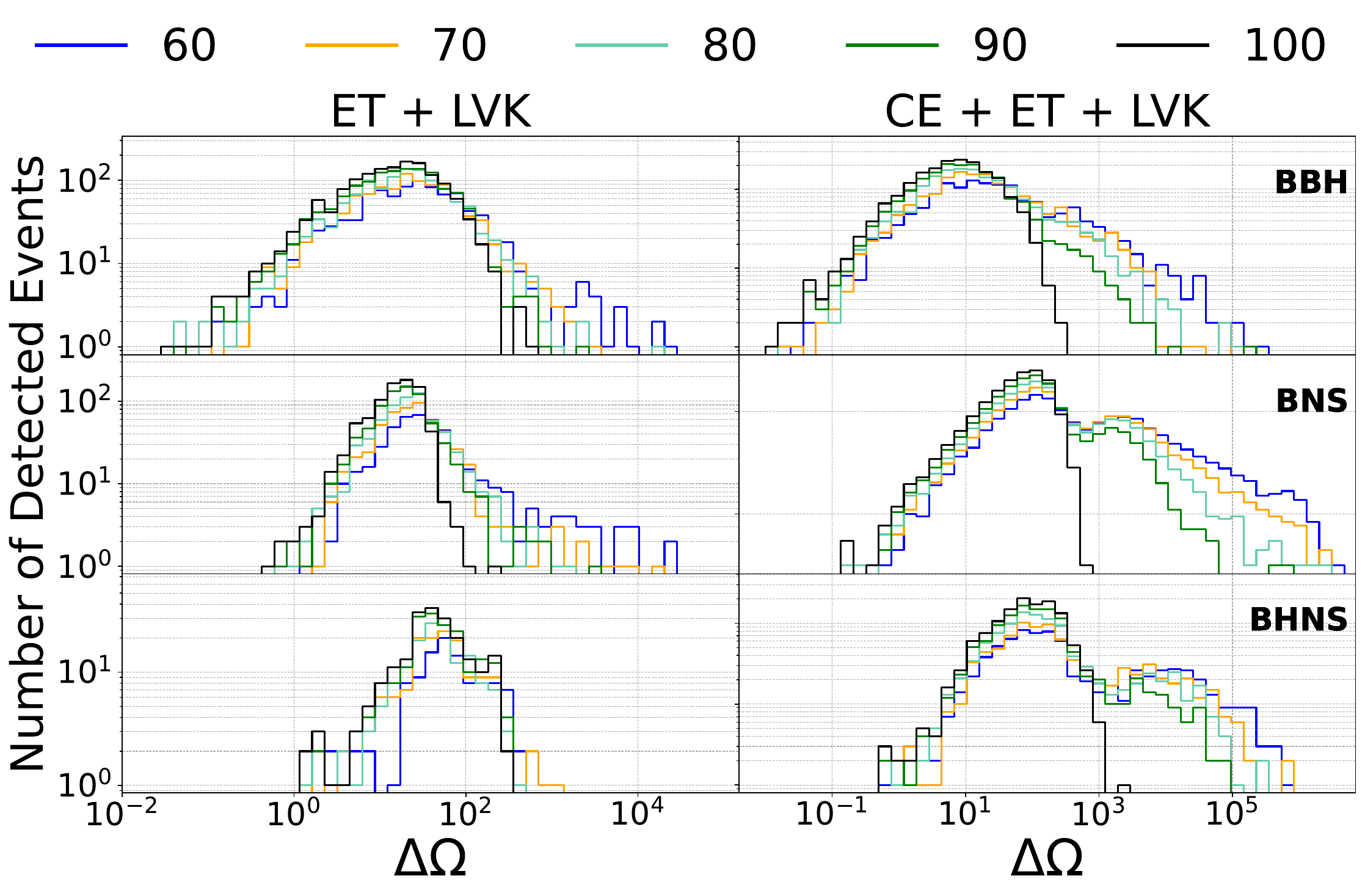}
\caption{
Impact of realistic values of the duty cycle.}
\label{fig:DCs}
\end{figure}

As discussed in Section~\ref{sec:dc}, gravitational-wave detectors do not operate continuously, and it is important to quantify the impact of a non-ideal duty cycle on observational outcomes.
We find that reduced duty cycles have a limited effect on the detection efficiency of 3G networks, primarily impacting the (unobserved) low-SNR tail of the distribution, as illustrated in Figure~\ref{fig:DCs}, which highlights the robustness of 3G configurations to variations in duty cycle. The impact on the estimation of chirp mass and symmetric mass ratio is minor. However, the most significant degradation is observed in the uncertainties of the luminosity distance and sky localization, particularly in configurations where CE and ET operate in conjunction with the LVK network.
This reduction in precision may affect both dark and bright siren cosmology, as fewer events meet the criteria for accurate host galaxy association or electromagnetic follow-up.

\subsection{LSST kilonova detections}
\label{res:kn}

Here, for the gravitational-wave detections, we adopt the CE+ET+LVK network configuration with a 70\% duty cycle. We use the waveform models \texttt{IMRPhenomXPHM} for BHNS systems and \texttt{IMRPhenomXHM} for BNS systems.
For the electromagnetic counterparts, we adopt the $5\sigma$ limiting magnitudes corresponding to 180-second exposures with the Rubin Observatory, as reported in \citet[Table1]{Andreoni:2024pkp} and in Table~\ref{tab:em_gw_detections} (in parentheses), which are representative of Target-of-Opportunity follow-up observations. While future follow-up capabilities may vary, these limits provide a realistic estimate of the performance expected from next-generation facilities available at the time the CE+ET+LVK network becomes operational.
See \citet{Colombo:2023une,Colombo:2025sdm,Loffredo:2024gmx} for a discussion that includes the modeling of kilonova light-curve.

\begin{figure}
\centering
\includegraphics[trim={0 0 0 0}, clip, width=1\columnwidth]{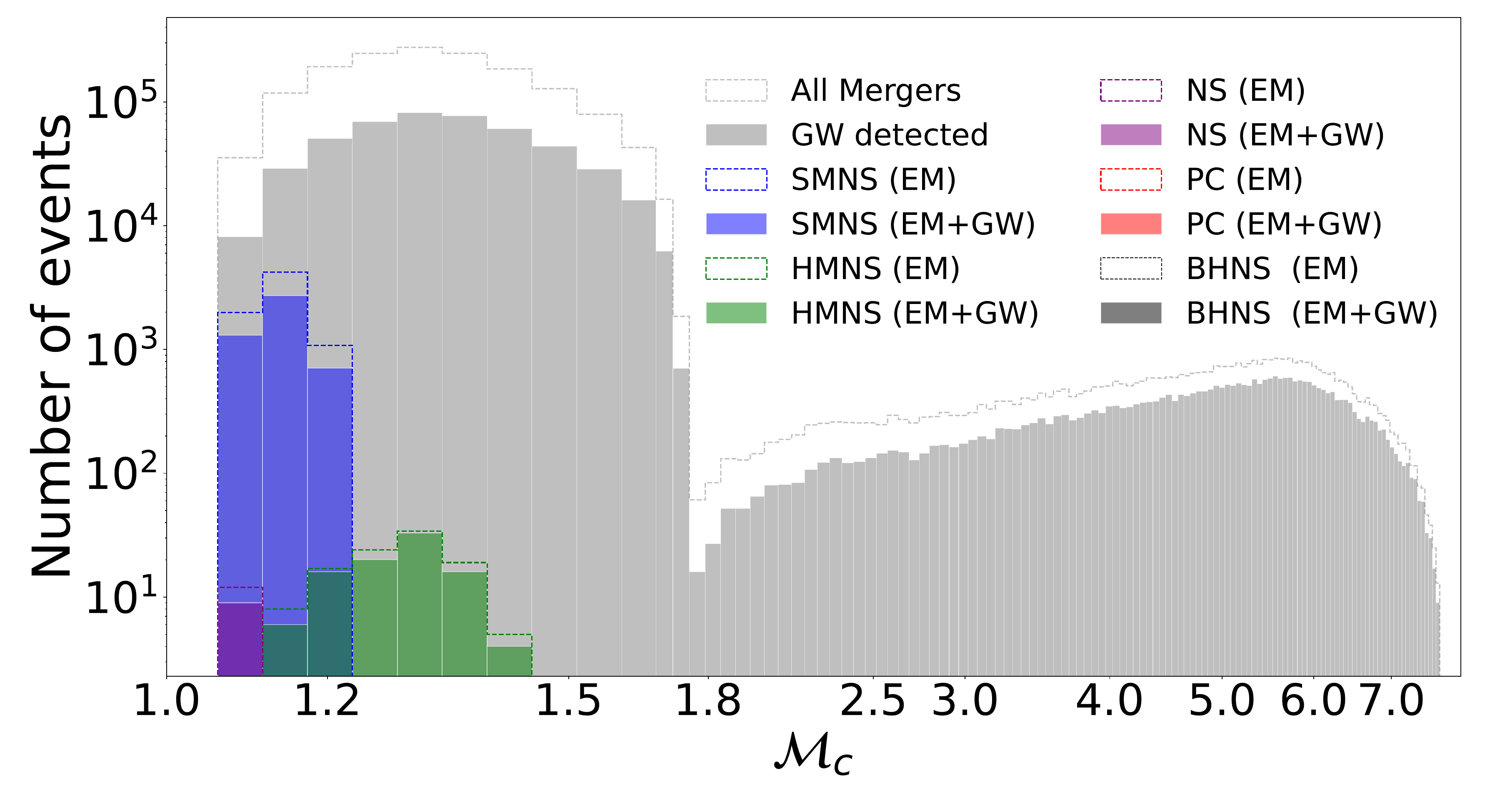}
\caption{  Chirp mass distribution of kilonovae detected by LSST in any band, categorized by merger remnant type. Dashed (empty) histograms correspond to all detected kilonovae, while solid (filled) histograms indicate those with associated gravitational-wave counterparts. The gray bars in the background represent the total merger population of BNS and BHNS systems (light gray) and GW events detected by the CE+ET+LVK network (dark gray). See Table~\ref{tab:em_gw_detections}.}
\label{fig:Apparent_Mag-types}
\end{figure}

Figure~\ref{fig:Apparent_Mag-types} presents the chirp mass distribution of kilonovae detected by LSST in any band, classified by merger remnant type. The filled bars represent events with a gravitational-wave counterpart, while the dashed outlines include all detected kilonovae. Table~\ref{tab:em_gw_detections} provides the number of electromagnetic detections per filter and remnant type, with and without associated gravitational-wave signals, along with the total detections across the full survey area of 15,833~deg$^2$ (36 tiles).

We find that the majority of detected kilonovae will originate from mergers resulting in supermassive neutron star remnants. The relatively small fraction of mergers with detectable electromagnetic counterparts arises because more massive systems, which are gravitationally easier to detect, tend to produce fainter kilonovae. Conversely, mergers resulting in brighter kilonovae generally involve lower-mass systems, which are gravitationally harder to detect.
From Figure~\ref{fig:Apparent_Mag-types} and Table~\ref{tab:em_gw_detections}, we conclude that significant improvements in electromagnetic follow-up capabilities are required to increase the number of detectable kilonovae.

Note that all events are included here, and those with low SNR are generally harder to localize. 
The Rubin Observatory has a field of view of 9.6~deg$^2$, which poses a challenge for identifying the kilonova counterpart of a neutron star merger with $\Delta \Omega > 100~\text{deg}^2$.
From Figure~\ref{DL-omega_err_net}, we find that 85\% of the BNS mergers satisfy $\Delta \Omega < 100~\text{deg}^2$.
See \citet{Menote:2025dew} for the forecasted cosmological constraints from the bright sirens in the \texttt{CosmoDC2\_BCO} catalog, obtained for different cuts in sky-localization precision.

\begin{table}
\centering
\caption{Electromagnetic detections by LSST (regular) and associated gravitational-wave counterparts detected by CE+ET+LVK (bold), per photometric band (limiting magnitude in parentheses).
Colors indicate the LSST filters. The assumed survey parameters are: 15\,833 deg$^2$ coverage (36 tiles), 10-year duration, and 70\% duty cycle.
\label{tab:em_gw_detections}}
\setlength{\tabcolsep}{9pt}
\renewcommand{\arraystretch}{1.3}
\begin{tabular}{l c c c c c c}
\toprule
Filter & NS & SMNS & HMNS & PC & BHNS  \\
\midrule
\rowcolor{uBand!25} \( u\,(24.9) \) & 1/\textbf{1} & 295/\textbf{259} & 0/\textbf{0} & 0/\textbf{0} & 0/\textbf{0} \\
\rowcolor{gBand!25} \( g \,(26.0) \) & 11/\textbf{8} & 7212/\textbf{4677} & 92/\textbf{80} & 1/\textbf{1} & 0/\textbf{0} \\
\rowcolor{rBand!25} \( r\, (25.7) \) & 9/\textbf{8} & 4841/\textbf{3395} & 89/\textbf{80} & 2/\textbf{2} & 0/\textbf{0} \\
\rowcolor{iBand!25} \( i\,(25.0) \) & 4/\textbf{4} & 1467/\textbf{1177} & 37/\textbf{36} & 1/\textbf{1} & 0/\textbf{0} \\
\rowcolor{zBand!25} \( z\,(24.3) \) & 2/\textbf{2} & 283/\textbf{249} & 6/\textbf{6} & 0/\textbf{0} & 0/\textbf{0} \\
\rowcolor{yBand!25} \( y\,(23.1) \) & 0/\textbf{0} & 7/\textbf{6} & 0/\textbf{0} & 0/\textbf{0} & 0/\textbf{0} \\
\rowcolor{white} Total & 12/\textbf{9} & 7297/\textbf{4727} & 107/\textbf{95} & 2/\textbf{2} & 0/\textbf{0} \\
\bottomrule
\end{tabular}
\end{table}

\section{Conclusions} \label{conclusions}

We presented a comprehensive set of synthetic catalogs of gravitational-wave and electromagnetic counterparts for binary compact object mergers, built through a consistent pipeline that associates GW events with realistic host galaxies from the \texttt{CosmoDC2} catalog. These catalogs support both dark and bright siren cosmology, enabling detailed multimessenger analyses with future gravitational-wave detectors and LSST follow-up observations.

We systematically explored various detector network configurations, waveform models, and duty cycle assumptions to evaluate their impact on event detectability and parameter estimation accuracy. Our main findings include:
\begin{itemize}
\item Third-generation detectors, notably the Einstein Telescope and Cosmic Explorer, substantially increase the detection rate and significantly improve parameter estimation, particularly for luminosity distance and sky localization.

\item Second-generation detectors remain essential for precise sky localization and distance estimation, especially for neutron star mergers, even when operated alongside 3G observatories. With LVK active, the triangular ET configuration and alternative 2ET-L configuration yield comparable performance.

\item Parameter uncertainties improve significantly when higher-order modes are included in the waveform modeling.

\item Assuming a simplified Target-of-Opportunity strategy, approximately 5000 kilonovae with gravitational-wave counterparts could be detectable by an LSST-like survey over a 10-year period, covering a 16000~deg$^2$ footprint in coordination with the CE+ET+LVK network operating at 70\% duty cycle. These events are expected to arise primarily from low-mass binary neutron star mergers forming supermassive neutron star remnants. However, this number represents only a small fraction of the total neutron star mergers detectable by third-generation gravitational-wave detectors. Given current LSST ToO capabilities, many kilonovae would remain undetected, highlighting the need for deeper, wide-field follow-up observations or complementary facilities. 
These projections rely on several simplifying assumptions—such as the adopted NSNS merger rate, the kilonova luminosity distribution, and the scheduling and sensitivity of future surveys—which introduce significant uncertainties. As such, the estimated detection numbers should be interpreted with caution.
\end{itemize}

The outcomes from this study form the publicly accessible \texttt{CosmoDC2\_BCO} database, available at \href{https://github.com/LSSTDESC/CosmoDC2_BCO}{github.com/LSSTDESC/CosmoDC2\_BCO}. A distinctive feature of this dataset is the coherent association of GW and EM signals with detailed host galaxy information from the \texttt{CosmoDC2} catalog. Each entry shares a unique \texttt{Galaxy\_ID}, facilitating straightforward crossmatching with extensive galaxy properties and observational systematics. The database comprises three primary catalogs:
\begin{itemize}
\item The \textit{Input Catalog} assigns intrinsic and extrinsic GW parameters to each host galaxy.
\item The \textit{Output Catalog} provides network SNRs and parameter covariance matrices for detected events, available across different detector configurations, waveform models, and duty cycles.
\item The \textit{Kilonova Value-Added Catalog} includes detailed classification, luminosities, and photometric information relevant for electromagnetic follow-up.
\end{itemize}

Future updates will focus on improved merger-rate modeling based on updated simulations within our ab initio framework, enhanced host-galaxy weighting schemes that incorporate luminosity and star-formation information, integration with next-generation LSST galaxy catalogs, more advanced inference methods based on the DALI approximation, and a more realistic electromagnetic follow-up strategy \citep{Colombo:2023une,Colombo:2025sdm,Loffredo:2024gmx}.

\section*{Acknowledgements}

This paper has undergone internal review in the LSST Dark Energy Science Collaboration.
RM developed the methodology, contributed to the interpretation, wrote the first draft, developed the code, and produced the data products.	
VM designed the study, developed the methodology, contributed to the interpretation, and wrote the first draft.
RS contributed to the development of the methodology and to the interpretation.
FAO and CRB contributed to the critical revision of the manuscript.\\
We thank Josiel de Souza for insightful discussions and support in using the \texttt{GWDALI} code, Francesco Iacovelli for clarifications regarding previous work, and Brian Metzger, Ben Margalit, and Luca Casagrande for discussions on kilonova properties. We are also grateful to Filippo Santoliquido for sharing the merger rate models.
We acknowledge Geraint Pratten for helpful input on waveform approximants and Joshua Smith for discussions concerning the CE site. We also thank Luca Amendola, Jaziel Coelho, Eleonora Loffredo, Michela Mapelli, Miguel Quartin, and the active DESC Slack community for valuable exchanges.\\
RM acknowledges financial support from FAPES (Brazil) and CAPES (Brazil) and thanks the Institute for Theoretical Physics at the University of Heidelberg for hosting his stay in Heidelberg. VM acknowledges partial support from CNPq (Brazil) and FAPES (Brazil).
RS acknowledges FAPESP grants n.\ 2022/06350-2 and 2021/14335-0, as well as CNPq grant n.\ 310165/2021-0
FAO acknowledges the SNSF under the grant 10.002.981.
CRB acknowledges the financial support from CNPq (316072/2021-4) and from FAPERJ (grants 201.456/2022 and 210.330/2022) and the FINEP contract 01.22.0505.00 (ref.\ 1891/22)\\
The DESC acknowledges ongoing support from the Institut National de 
Physique Nucl\'eaire et de Physique des Particules in France; the 
Science \& Technology Facilities Council in the United Kingdom; and the
Department of Energy and the LSST Discovery Alliance
in the United States.  DESC uses resources of the IN2P3 
Computing Center (CC-IN2P3--Lyon/Villeurbanne - France) funded by the 
Centre National de la Recherche Scientifique; the National Energy 
Research Scientific Computing Center, a DOE Office of Science User 
Facility supported by the Office of Science of the U.S.\ Department of
Energy under Contract No.\ DE-AC02-05CH11231; STFC DiRAC HPC Facilities, 
funded by UK BEIS National E-infrastructure capital grants; and the UK 
particle physics grid, supported by the GridPP Collaboration.  This 
work was performed in part under DOE Contract DE-AC02-76SF00515.
\\
The authors acknowledge the use of computational resources from the \href{https://computacaocientifica.ufes.br/scicom}{Sci-Com Lab} of the Department of Physics at UFES, supported by FAPES, CAPES, and CNPq.

\section*{Data availability}

The data underlying this article will be shared on reasonable request to the corresponding author.

\bibliographystyle{mnrasArxiv}
\bibliography{biblio}

\appendix

\section{Fisher validation}
\label{ap:validation}

The Fisher matrix provides a useful approximation of the likelihood near its maximum but requires validation to quantify its accuracy and biases. We thus conducted Markov Chain Monte Carlo (MCMC) analyses for approximately 300 randomly selected BBH mergers using the \texttt{IMRPhenomXPHM} waveform, considering both LVK and 2CE+ET networks at a 100\% duty cycle.

\begin{figure}
\centering 
\includegraphics[trim={0 0 0 0}, clip, width= 1 \columnwidth, height =7cm]{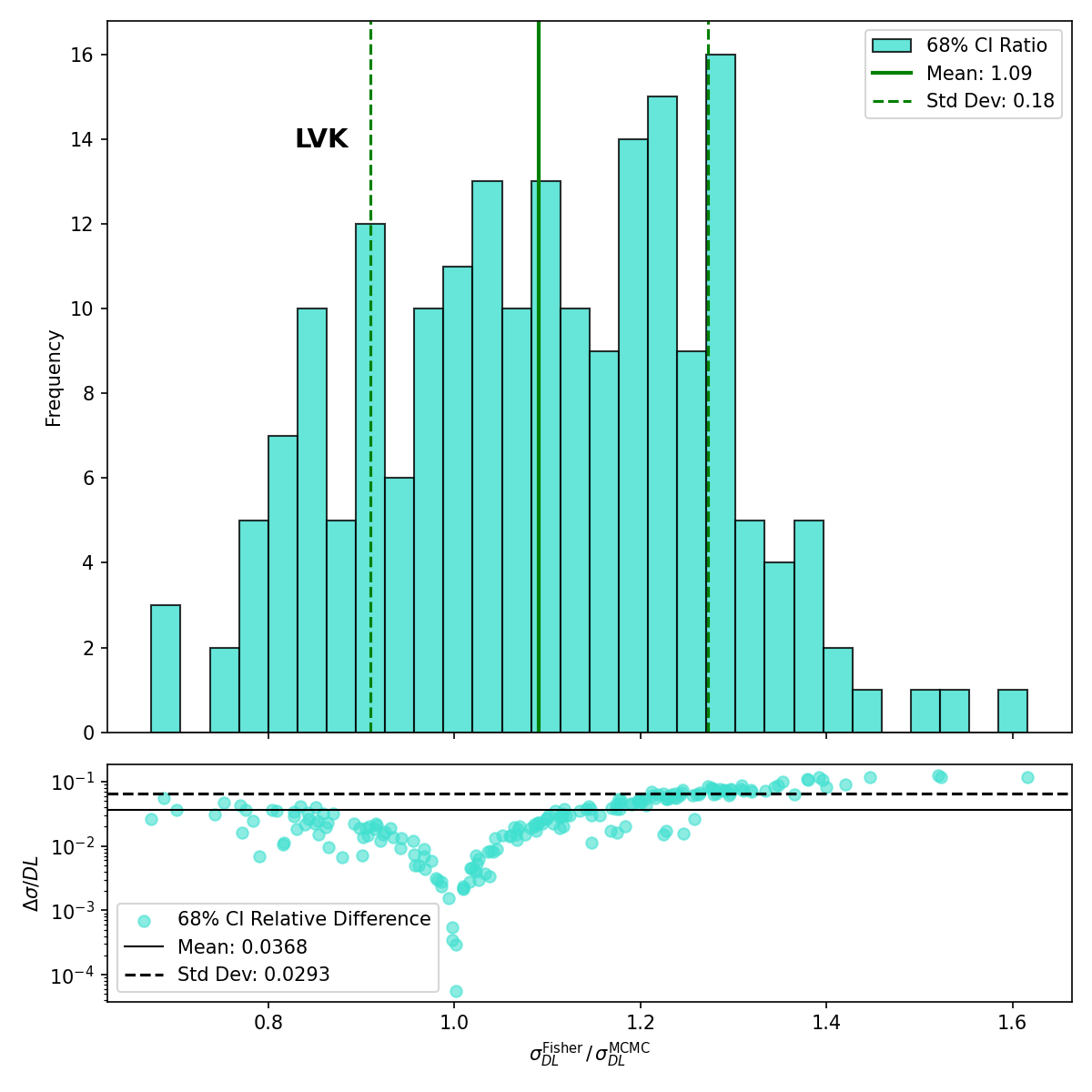}
\includegraphics[trim={0 0 0 0}, clip, width= 1 \columnwidth, height =7cm]{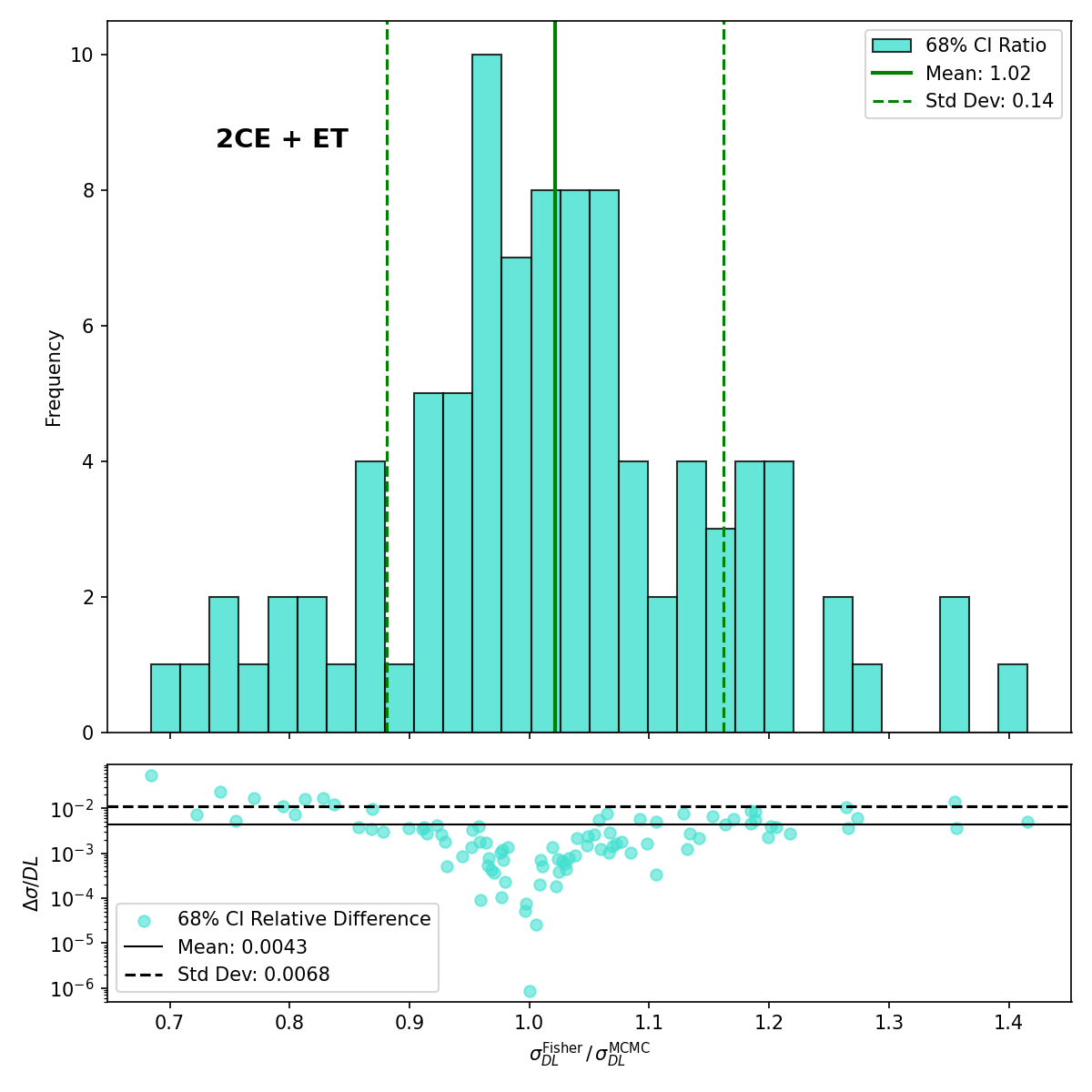}
\caption{Distribution of the ratios between the 68\% uncertainties estimated via the Fisher matrix and those obtained from MCMC sampling for the LVK (top) and 2CE+ET (bottom) networks with a 100\% duty cycle.\label{ratio_2CEET_and_LVK}}
\end{figure}

We found that fully uninformative priors on the 15 GW parameters significantly hindered convergence using the default \texttt{bilby} sampler, preventing chains from reaching a steady state. To address this, we imposed uniform priors centered at the fiducial parameter values with widths of $\pm 5\sigma$, where $\sigma$ denotes the marginalized uncertainty obtained from the Fisher matrix.
Figure~\ref{ratio_2CEET_and_LVK} presents distributions of the ratio between the 68\% credible uncertainties estimated by the Fisher matrix and those derived from the MCMC posterior samples.
We find that the Fisher matrix tends to slightly overestimate uncertainties. For LVK, the mean ratio is $1.09 \pm 0.18$, while for 2CE+ET the mean is $1.02 \pm 0.14$. As expected, the latter is closer to unity because the higher network sensitivity yields more Gaussian posterior distributions.

\section{Effect of approximant}
\label{ap:approximant}

\begin{figure}
\centering 
\includegraphics[trim={0 0 0 0}, clip, width= 1 \columnwidth]{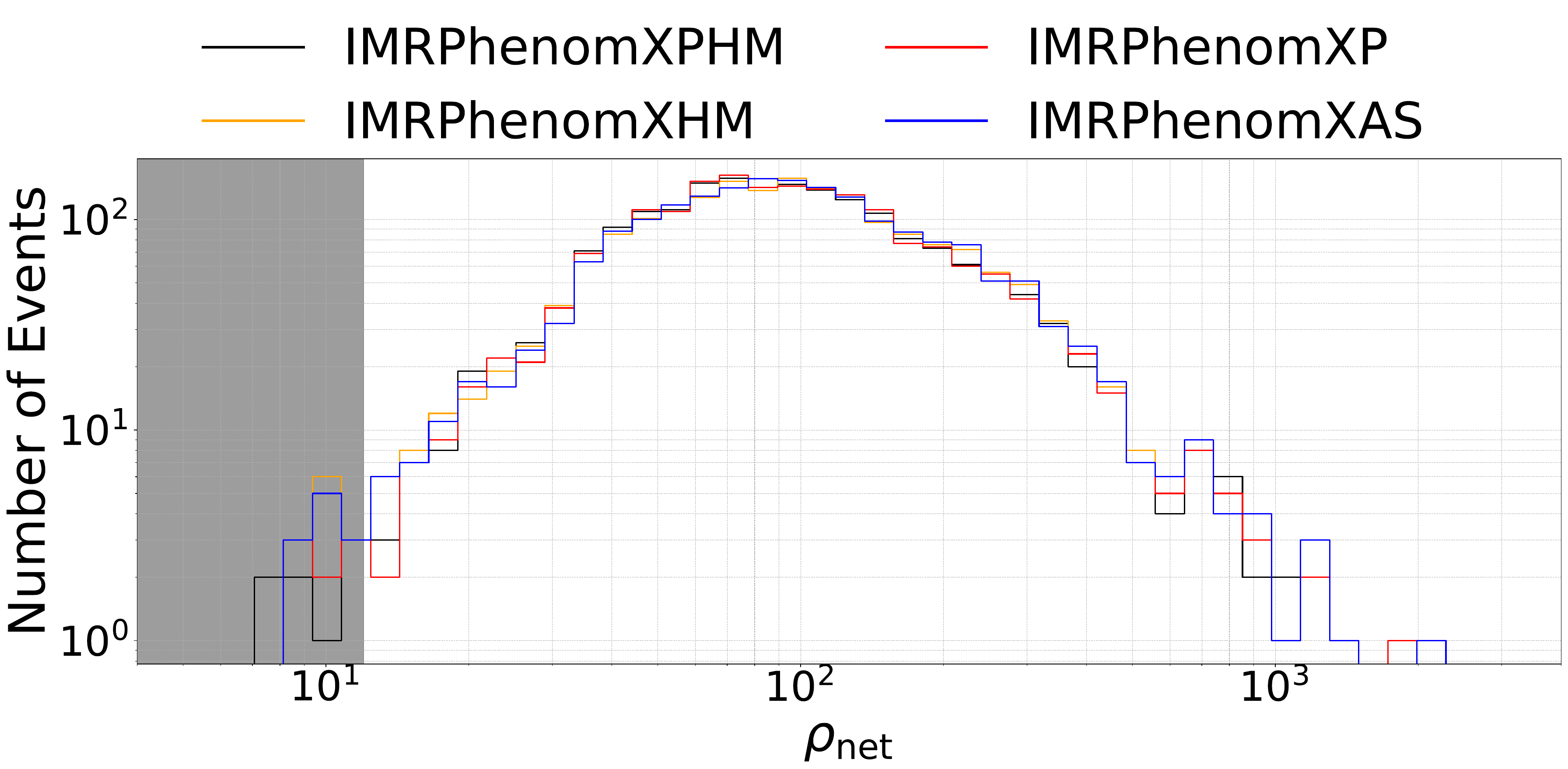}
\includegraphics[trim={0 0 0 0}, clip, width= 1 \columnwidth]{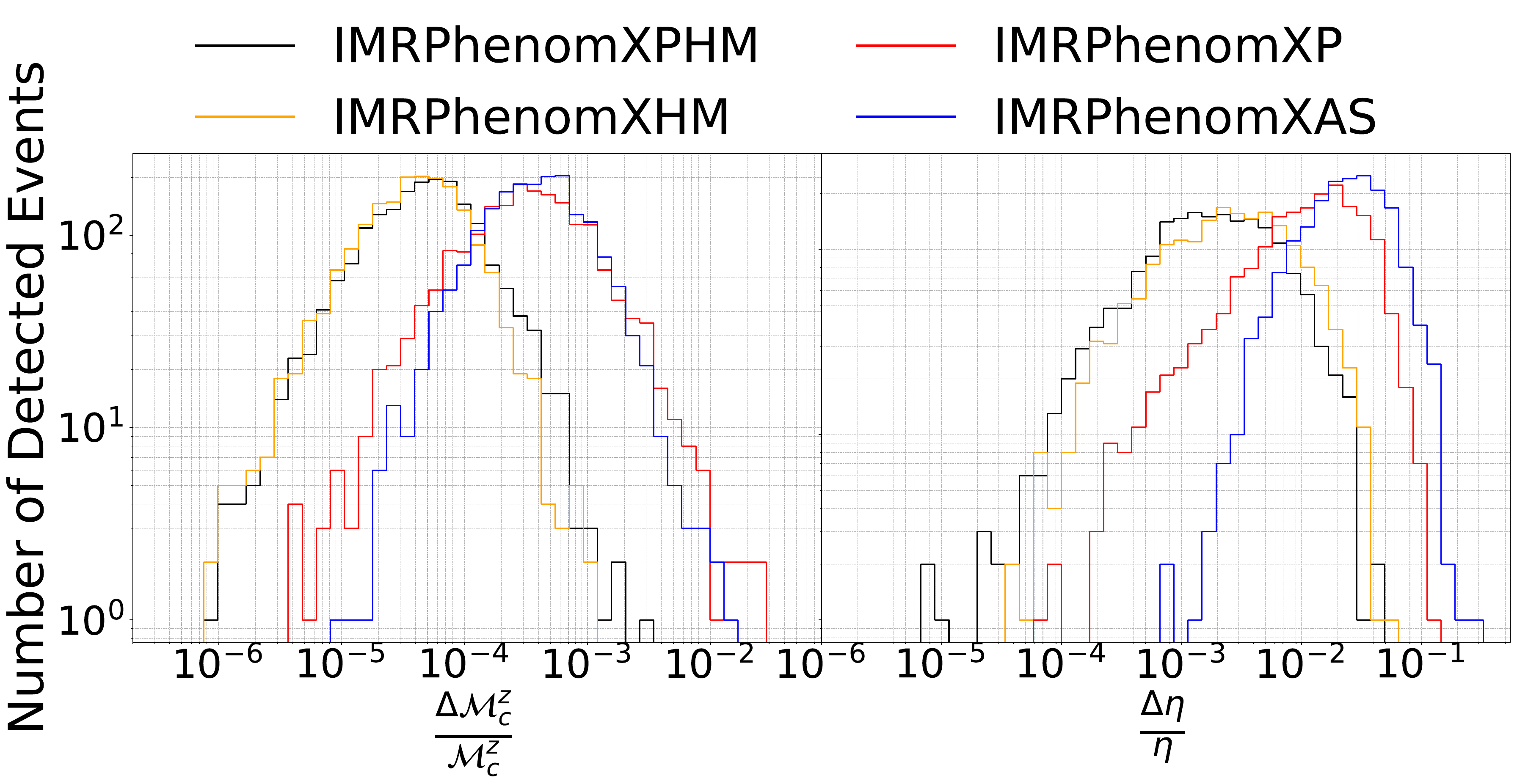}
\includegraphics[trim={0 0 0 0}, clip, width= 1 \columnwidth]{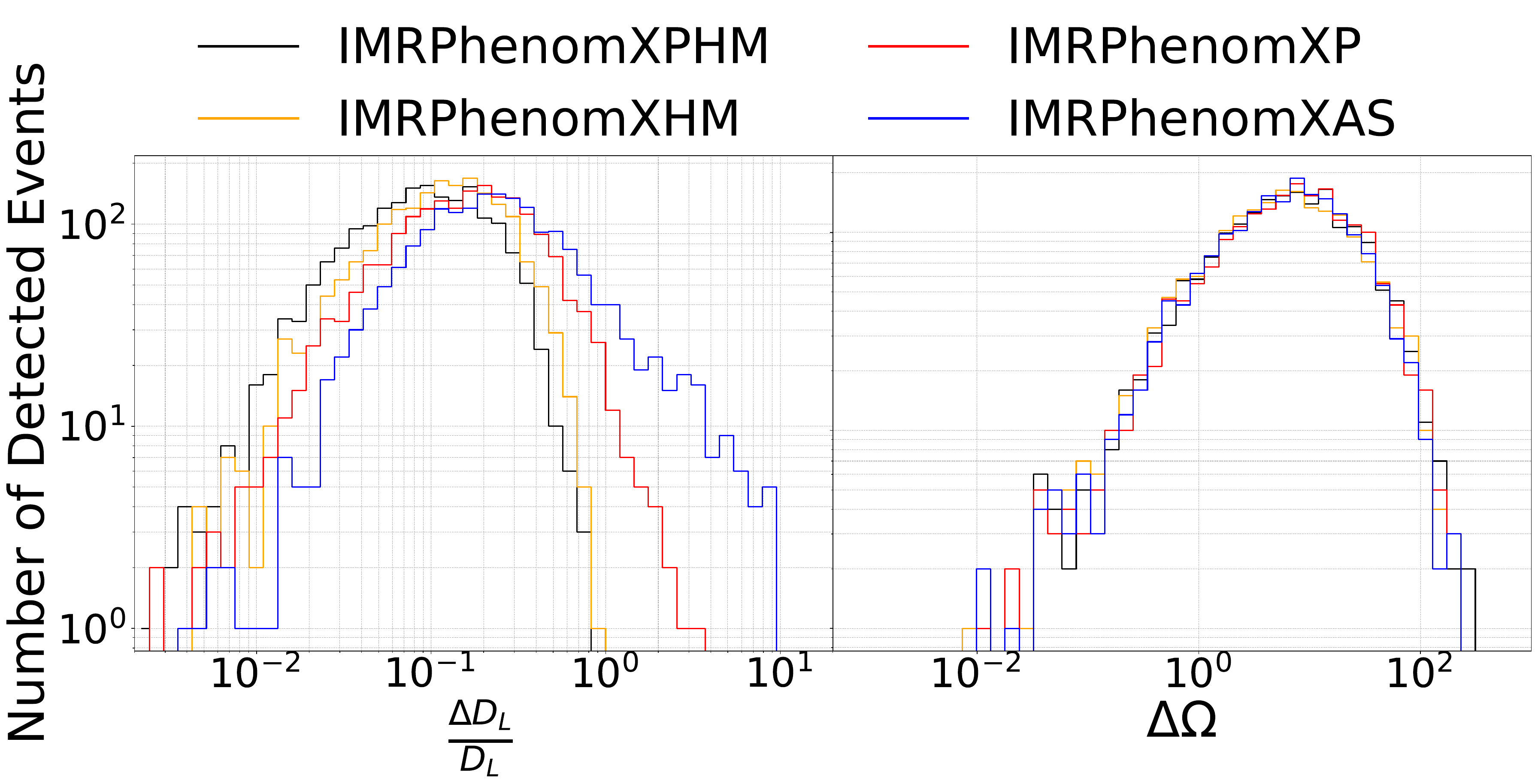}
\caption{Impact of waveform model on observable quantities for BBH mergers detected by the CE+ET+LVK network with 100\%\ duty cycle. }
\label{fig:approximant}
\end{figure}

Here we consider the CE+ET+LVK network with 100\% duty cycle and assess the impact of different waveform models.  
Figure~\ref{fig:approximant} shows that the choice of waveform has little effect on the SNR distribution for BBH mergers. However, it has a pronounced impact on the precision of the chirp mass and symmetric mass ratio.
In particular, incorporating higher-order modes significantly improves mass measurements, as these modes enhance waveform accuracy without introducing additional parameters. In contrast, including spin precession increases the dimensionality of the parameter space by adding four spin orientation variables, which can amplify degeneracies and thus does not necessarily result in smaller uncertainties.

An interesting improvement is also observed for the luminosity distance, driven by tighter constraints on the chirp mass and the partial breaking of the $D_L$--$\iota$ degeneracy enabled by higher-order modes. In contrast, sky localization remains largely unaffected by the choice of waveform model. These findings are consistent with \citet{Liu:2024jkj}.

\section{Further analyses}
\label{ap:further}

\begin{figure}
\centering 
\includegraphics[trim={0 0 0 0}, clip, width= 1 \columnwidth]{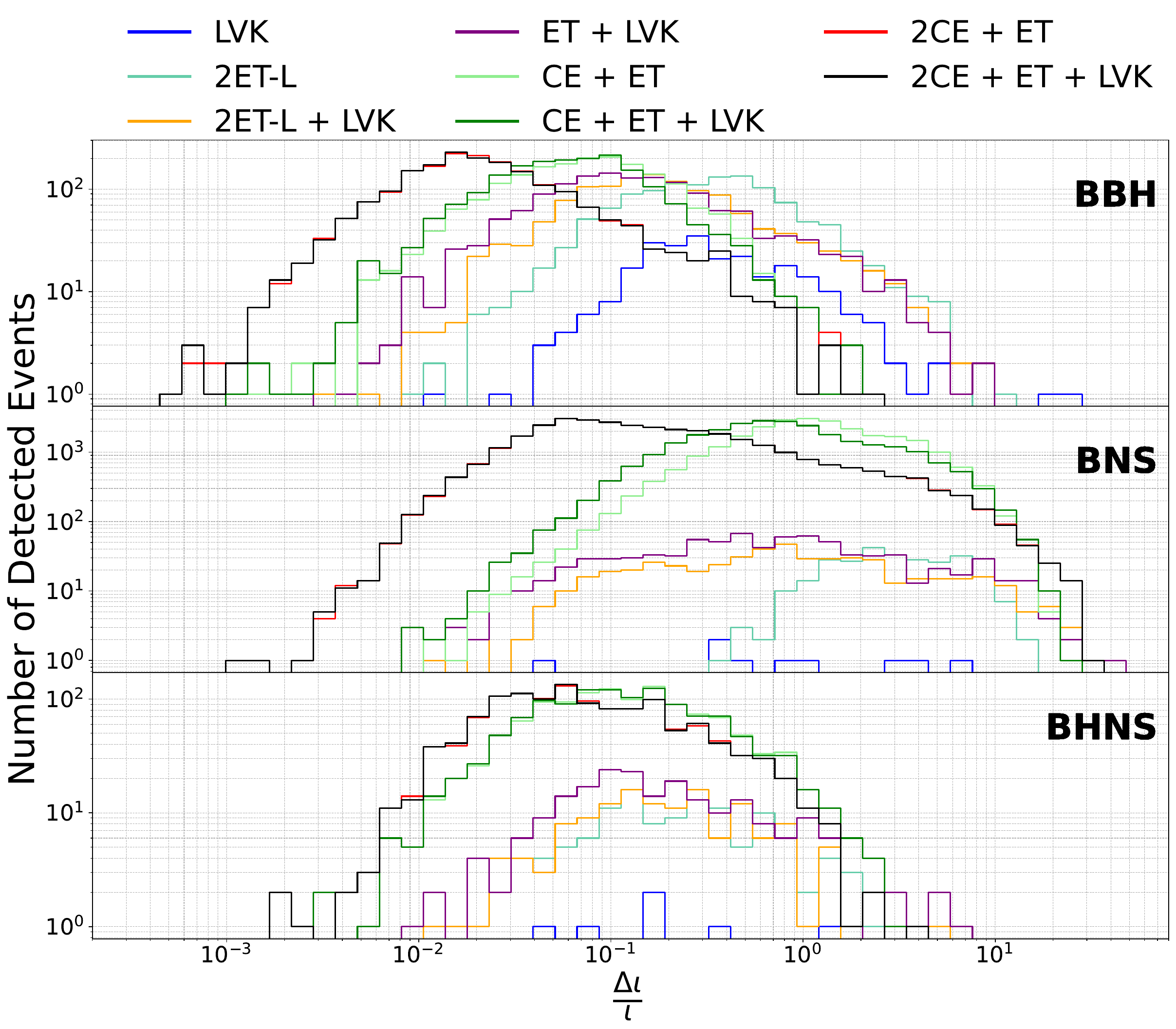}
\caption{Distributions of BBH (top), BNS (middle), and BHNS (bottom) merger uncertainties on the inclination angle $\iota$ from \texttt{tile~\#0}. Results assume a 100\%\ duty cycle. }
\label{fig:iota_uncertainties}
\end{figure}

Figure~\ref{fig:iota_uncertainties} shows the distributions of the uncertainty on the inclination angle $\iota$ for BBH, BNS, and BHNS mergers from \texttt{tile~\#0} and illustrates the progressive enhancement in $\iota$ precision as additional detectors are included in the network. This improvement is particularly relevant for cosmology, as the luminosity distance is strongly degenerate with the inclination angle. The inclusion of 3G detectors enables the detection of a large number of events with highly constrained $\iota$, although a non-negligible tail of high-uncertainty events persists, mostly associated with low-SNR detections.
The most significant gains are observed with networks that include three 3G detectors, especially for BBH and BNS populations.
Further plots and analysis are available at \href{https://github.com/LSSTDESC/CosmoDC2_BCO}{github.com/LSSTDESC/CosmoDC2\_BCO}, including cumulative distributions, sky maps and distributions of the uncertainties on all the intrinsic and extrinsic parameters.

\label{lastpage}
\end{document}